\documentclass[preprint,authoryear,12pt]{elsarticle}
\usepackage{graphicx}
\def\figfigincl#1#2#3{\includegraphics[width=#1]{figures/#2.eps}%
    \caption{\small #3}\label{fig:#2}}

\journal{Icarus}
\begin{document}

\title{Asteroid family ages}
\author[pi,sds]{Federica Spoto} 
\ead{spoto@spacedys.com}
\author[pi]{Andrea Milani}
\author[bg]{Zoran Kne\v zevi\'c}
\address[pi]{Dipartimento di Matematica, Universit\`a di Pisa,
        Largo Pontecorvo 5,
        56127 Pisa, Italy}
\address[bg]{Astronomical Observatory, Volgina 7,
         11060 Belgrade 38, Serbia}
\address[sds]{SpaceDyS srl, Via Mario Giuntini 63, 
              56023 Navacchio di Cascina, Italy}

\begin{abstract}

A new family classification, based on a catalog of proper elements
with $\sim 384,000$ numbered asteroids and on new methods is
available. For the $45$ dynamical families with $>250$ members
identified in this classification, we present an attempt to obtain
statistically significant ages: we succeeded in computing ages for
$37$ collisional families.

We used a rigorous method, including a least squares fit of the two
sides of a V-shape plot in the proper semimajor axis, inverse
diameter plane to determine the corresponding slopes, an advanced
error model for the uncertainties of asteroid diameters, an iterative
outlier rejection scheme and quality control. The best available
Yarkovsky measurement was used to estimate a calibration of the
Yarkovsky effect for each family.
The results are presented separately for the families originated in
fragmentation or cratering events, for the young, compact families and
for the truncated, one-sided families. For all the computed ages the
corresponding uncertainties are provided, and the results are
discussed and compared with the literature. The ages of several
families have been estimated for the first time, in other cases
the accuracy has been improved.
We have been quite successful in computing ages for old families, we
have significant results for both young and ancient, while we have
little, if any, evidence for primordial families. We found 2 cases
where two separate dynamical families form together a single V-shape
with compatible slopes, thus indicating a single collisional event. We
have also found 3 examples of dynamical families containing multiple
collisional families, plus a dubious case: for these we have obtained
discordant slopes for the two sides of the V-shape, resulting in
distinct ages. We have found 2 cases of families containing a
conspicuous subfamily, such that it is possible to measure the slope
of a distinct V-shape, thus the age of the secondary collision. We
also provide data on the central gaps appearing in some families.
The ages computed in this paper are obtained with a single and uniform
methodology, thus the ages of different families can be compared,
providing a first example of collisional chronology of the asteroid
main belt.

\end{abstract}

\maketitle


\section{Introduction}
\label{sec:intro}

One of the main purposes for collecting large datasets on asteroid
families is to constrain their ages, that is the epoch of the impact
event generating a collisional family. A collisional family not always
coincides with the dynamical family detected by density contrast in 
the proper elements space. More complicated cases occur, such as a
dynamical family to be decomposed in two collisional families,
or the opposite case in which a collisional family is split in
two density contrast regions by some dynamical instability.

Although other methods are possible, currently the most precise method
to constrain the age of a collisional family (for the ages
older than $\sim 10$ My) exploits non-gravitational perturbations,
mostly the Yarkovsky effect~\citep{vok_2000}. These effects generate
secular perturbations in the proper elements of an asteroid which are
affected not just by the position in phase space, but also by the
Area/Mass ratio, which is inversely proportional to the asteroid
diameter $D$. Thus, the main requirements are to have a list of family
members with a wide range of values in $D$, enough to detect the
differential effect in the secular drift of the proper elements
affecting the shape of the family, and to have a large enough
membership, to obtain statistically significant results.

Recently \cite{bigdata} have published a new family classification by
using a large catalog of proper elements (with $>330,000$
numbered asteroids) and with a classification method improved
  with respect to past methods. This method is an extension of the
Hierarchical Clustering Method (HCM)~\citep{zapetal90}, with special
provisions to be more efficient in including large numbers of small
objects, while escaping the phenomenon of chaining. Moreover, the new
method includes a feature allowing to (almost) automatically update
the classification when new asteroids are numbered and their proper
elements have been computed. This has already been applied to extend
the classification to a source catalog with $\sim 384,000$ proper
elements, obtaining a total of $\sim 97,400$ family members
\citep{namur_update}. In this paper we are going to use the
classification of \cite{bigdata}, as updated by \cite{namur_update},
and the data are presently available on
AstDyS\footnote{http://hamilton.dm.unipi.it/astdys/index.php?pc=5}.

This updated classification has $21$ dynamical families with $>1,000$
members and another $24$ with $>250$ members. The goal of this paper
can be simply stated as to obtain statistically significant age
constraints for the majority of these $45$ families.
Computing the ages for all would not be a realistic goal because there
are several difficulties. Some families have a very complex structure,
for which it is difficult to formulate a model, even with more than
one collision: these cases have required or need dedicated
studies. Some families are affected by particular dynamical
conditions, such as orbital resonances with the planets,
which result in more complex secular perturbations: these shall be the
subject of continuing work. The results for families with only a
moderate number of members (such as $250 - 300$) might have a low
statistical significance.

The age estimation includes several sources of uncertainty which
cannot be ignored. The first source appears in the formal accuracy in
the least square fit used in our family shape estimation methods.  The
uncertainty depends upon the noise resulting mostly from the
inaccuracy of the estimation of $D$ from the absolute magnitude
$H$. The second source of error occurs in the conversion of the
inverse slope of the family boundaries into age, requiring a Yarkovsky
calibration: this is fundamentally a relative uncertainty, and in most
cases it represents the largest source of uncertainty in the inferred
ages. In Sec.~\ref{s:yarko_cal} we give an estimate of this
uncertainty between $20\%$ and $30\%$.

As a result of the current large relative uncertainty of the
calibration, we expect that this part of the work will be soon
improved, thanks to the availability of new data. Thus the main result
of this paper are the inverse slopes, because these are derived by
using a consistent methodology and based upon large and comparatively
accurate data set. Still we believe we have done a significant
progress with respect to the previous state of the art by estimating
$37$ collisional family ages, in many cases providing the first
rigorous age estimate, and in all cases providing an estimated
standard deviation. The work can continue to try and extend the
estimation to the cases which we have found challenging.

Since this paper summarizes a complex data processing, with output
needed to fully document our procedures but too large, we decided to
include only the minimum information required to support our analysis
and results. Supplementary material, including both tables and plots,
is available from the web site \texttt{
  http://hamilton.dm.unipi.it/astdys2/fam\_ages/}.

\section{Least squares fit of the V-shape}
\label{sec:fit}

Asteroids formed by the same collisional event take the form of a V in
the (proper $a$ - $1/D$) plane. The computation of the family ages can
be performed by using this V-shape plots if the family is old enough
and the Yarkovsky effect dominates the spread of proper $a$, as
explained in~\citep[Sec. 5.2]{bigdata}. The key idea is to compute the
diameter $D$ from the absolute magnitude $H$, assuming a common
geometric albedo $p_v$ for all the members of the family. The common
geometric albedo is the average value of the known WISE
albedos~\citep{WISE10, Mainzer_WISE} for the asteroids in the
  family. Then we use the least squares method to fit the data with
two straight lines, one for the low proper $a$ (IN side) and the other
for the high proper $a$ (OUT side), as in \citet{bigdata}, with an
improved outlier rejection procedure, see \citep{carpino03} and
Sec.~\ref{s:outlier_rejection}.

\subsection{Selection of the Fit Region}

\begin{table}[h!]
  \footnotesize
  \centering
  \caption{Fit region: family number and name, explanation of the
    choice, minimum value of proper $a$, minimum value of the diameter
    selected for the inner and the outer side.}
  \label{tab:fit_region}
  \medskip
  \begin{tabular}{|l|rlr|rlr|}
    \hline
number/          &cause &min       &min  &cause &max       & min \\
name             &      &proper $a$&D IN &      &proper $a$& D OUT\\
\hline
158 Koronis      &  5/2 &   2.82   & 7.69& 7/3  &   2.96   & 5.00\\
24  Themis       & 11/5 &   3.075  &25.00& 2/1  &   3.24   &16.67\\
847 Agnia        &  8/3 &   2.70   & 4.55& 5/2  &   2.82   & 6.67\\
3395 Jitka       &  FB  &   2.76   & 1.33& 5/2  &   2.82   & 1.33\\
1726 Hoffmeister & 3-1-1&   2.75   & 5.00& 5/2  &   2.82   & 4.00\\ 
668 Dora         & 3-1-1&   2.75   & 5.88& 5/2  &   2.815  & 8.33\\ 
434 Hungaria     &  5/1 &   1.87   & 1.25& 4/1  &   2.00   & 1.25\\ 
480 Hansa        &  3/1?&   2.54   & 5.00& FB   &   2.71   & 6.67\\ 
808 Merxia       &  8/3 &   2.7    & 2.50& FB   &   2.80   & 2.00\\
3330 Gantrisch   & FB?  &   3.13   & 6.67& 5-2-2&   3.17   & 6.67\\ 
10955 Harig      &  FB  &   2.67   & 1.43& FB   &   2.77   & 1.82\\ 
293 Brasilia     &  5/2 &   2.83   & 2.50& FB   &   2.88   & 2.00\\
569 Misa         &  FB  &   2.62   & 3.33& FB   &   2.70   & 3.33\\
15124 2000EZ$_{39}$&  FB  &   2.62   & 2.00& FB   &   2.70   & 2.50\\
1128 Astrid      &  FB? &   2.755  & 2.22& 5/2  &   2.82   & 2.22\\ 
845 Naema        &  FB  &   2.91   & 2.86& 7/3  &   2.96   & 5.00\\
\hline
4 Vesta           &  7/2 &   2.25   & 2.50& 3/1  &   2.50   & 2.94\\ 
15 Eunomia        &  3/1?&   2.52   & 5.00& 8/3  &   2.71   & 5.00\\ 
20 Massalia       & 10/3 &   2.33   & 1.00& 3/1  &   2.50   & 0.91\\ 
10 Hygiea         & 11/5 &   3.07   & 7.14& 2/1  &   3.25   & 7.69\\ 
31 Euphrosyne     & 11/5 &   3.07   & 6.67& 2/1  &   3.25   & 6.67\\ 
3 Juno            & FB?  &   2.62   & 2.00& 8/3  &   2.70   & 2.50\\
163 Erigone       & 10/3 &   2.33   & 2.50& 2/1M &   2.42   & 2.50\\ 
\hline
3815 K\"onig     & FB   &   2.56   & 2.20& FB   &   2.585  & 2.20\\ 
396 Aeolia       & FB   &   2.73   & 1.67& FB   &   2.755  & 2.00\\ 
606 Brangane     & FB   &   2.57   & 1.67& FB   &   2.595  & 1.67\\ 
1547 Nele        & FB   &   2.64   & 1.67& FB   &   2.648  & 1.67\\
18405 1993FY$_{12}$& FB  &   2.83   & 2.50& FB   &   2.85  &  2.00\\
\hline
170 Maria        & 3/1? &          &     &  FB  &   2.665  & 4.00\\    
93 Minerva       & FB   &   2.71   & 4.00&  5/2 &          &     \\ 
2076 Levin       & 7/2  &          &     &  FB  &   2.34   & 2.50\\ 
3827 Zdenekhorsky& 8/3  &   2.7    & 2.00& 1-1C &          &     \\ 
1658 Innes       & 3/1? &          &     & 11/4 &   2.645  & 2.00\\
375 Ursula       & FB   &   3.1    &12.50&  2/1 &          &     \\  
\hline
\end{tabular}
\end{table}

Most families are bounded on one side or on both sides by
resonances. Almost all these resonances are strong enough to eject
most of the family members that fell into the resonances into
unstable orbits. In these cases the sides of the V are cut by vertical
lines, that is by values of $a$, which correspond to the border of the
resonance. For each family we have selected the fit region taking into
account the resonances at the family boundaries. The fit of the slope
has to be done for values of $1/D$ below the intersection of one of
the sides of the V affected by the resonance and the resonance border
value of proper $a$. In Table~\ref{tab:fit_region} we report the
values for $a$ and $D$, and the cause of each selection.

The \textit{cause} of each cut in proper $a$ is a mean motion
resonance, in most cases a 2-body resonance with Jupiter, in few cases
either a 2-body resonance with Mars or a 3-body resonance with Jupiter
and Saturn. When no resonance with this role has been identified, we
use the label FB (for Family Box) to indicate that the family ends
where the HCM procedure does not anymore detect a significant density
contrast (with respect to the local background). This is affected by
the depletion of the proper elements catalog due to the completeness
limit of the surveys: the family may actually contain many smaller
asteroids beyond the box limits, but they have not been discovered
yet. On the contrary when the family range in proper $a$ is delimited
by strong resonances, the family members captured in them can be
transported far in proper $e$ (and to a lesser extent in proper
$\sin{I}$) to the point of not being recognizable as members; over
longer time spans, they can be transported to planet-crossing orbits
and removed from the main belt altogether.

The tables in this paper are sorted in the same way: there
are four parts, dedicated to families of the types fragmentation,
cratering, young, one-sided; inside each group the families are sorted
by decreasing number of members. In some cases the tables have been
split in four sub-tables, one for each type.

In two cases we have already defined the fit region in such a way that
we can include two families in a single V-shape. This \textit{family
  join} is justified later, in Section~\ref{s:results}, by showing
that the two dynamical families can be generated by a single
collision. This applies to the join of 10955 with 19466 and to the
join of 163 with 5206. Note that the join of two families, justified
by the possibility to fit together in a single V-shape with a common
age, is conceptually different from the \textit{merge} of two families
due to intersections, discussed in \citep{bigdata,namur_update};
however, the practical consequences are the same, namely one family is
included in another one and disappears from the list of families. 

For one-sided families we are also indicating the ``cause'' of the
missing side.  E.g., for 2076 the lack of the IN side of the V-shape
is due to the 7/2 resonance; on the other hand, the dynamical
family 883 could be the continuation of 2076 at proper $a$ lower than
the one of the resonance. However, the V-shape which would be obtained
by this join would have two very different slopes, thus it can be
excluded that they are the same collisional family.


For most families the ``cause'' of the delimitation in proper $a$, in
the sense above, can be clearly identified. However, some ambiguous
cases remain: e.g., for family 1128 the outer boundary could be due to
the 3-body resonance 3-1-1 (the three integer coefficients apply to
the mean motions of Jupiter, Saturn and the asteroid, respectively);
for family 3 the inner boundary could be due to 4-3-1. For family 3330
a 3-body resonance (not identified) at $a=3.129$ could be the cause of
the inner boundary.

For the one-sided family 3827 we do not know the cause of the missing
OUT side, although we suspect it has something to do with (1)
Ceres, given that the proper $a$ of Ceres is very close to the upper
limit of the family box.

The family of (3395) Jitka is a subfamily of the dynamical family
847. The family of (15124) 2000 EZ$_{39}$ is a subfamily of the
dynamical family 569. 

With 3/1? we are indicating 2 cases (480, 15) in which the
families could be delimited on the IN side by the 3/1 resonance (also
170, 1658 in which the 3/1 could be the cause of the missing IN side),
but the lower bound on proper $a$ appears too far from the Kirkwood
gap. This is a problem which needs to be investigated. 

\subsection{Binning and fit of the slopes}

Next we divide the $1/D$ axis into bins, as in
Figures~\ref{fig:20_bins_in} and~\ref{fig:20_bins_out}. The
partition is done in such a way that each bin contains roughly the
same number of members. 

The following points explain the main
  features of the method used to create the bins:
\begin{enumerate}
\item the maximum number of bins $N$ is selected for each
  family, depending upon the number of members of the family;
\item the maximum value of the standard deviation of the
  number of members in each bin is decided depending upon the number
  of members of the family;
\item the region between $0$ and the maximum value of $1/D$ is
  divided in $N$ bins;
\item the difference between the number of members in two
  consecutive bins is computed:
  \begin{itemize}
  \item[4.1] if the difference is less than the standard
    deviation, the bins are left as they are;
  \item[4.2] if the difference is greater than the standard
    deviation, the first bin is divided into smaller bins and then the
    same procedure is applied to the new bins.
  \end{itemize}
\end{enumerate}
This procedure is completely automatic, and it is the same both for
the inner and the outer side of a V-shape. In the example of the
Figures, namely the family of (20) Massalia, in the IN side there are
$84$ bins with a mean of $19$ members in each, with a STD of this
number $13$. In the OUT side there are $82$ bins with mean $19$ and
STD $11$.

In the case of the low $a$ side we select the minimum value of proper
$a$ and the corresponding $1/D$ in each bin, as in
Fig.~\ref{fig:20_bins_in}. For the other side we select the maximum
value of the proper semimajor axis and the corresponding $1/D$, as in
Fig.~\ref{fig:20_bins_out}. These are the data to be fit to determine
the slopes of the V-shapes: thus it is important to have enough bins
to properly cover the range in proper $a$.

\begin{figure}[h!]
\figfigincl{12 cm}{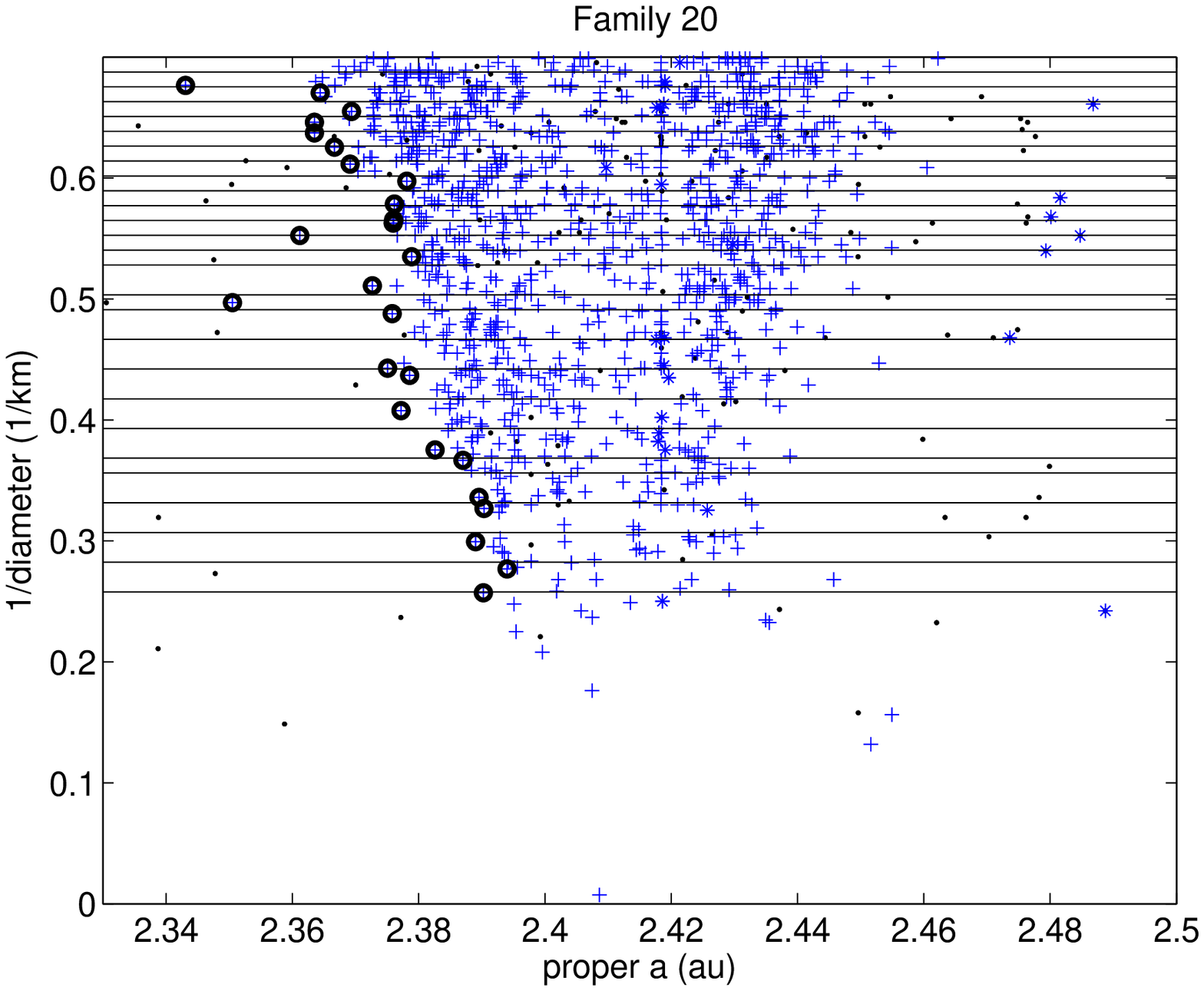}{Blow up of the bins for the
    inner side of the family of (20) Massalia. Crosses are the members
    of the family, points are background asteroids, stars are affected
    by the resonances. Circles are members of the family of (20)
    Massalia with the minimum value of proper $a$ and the
    corresponding $1/D$ in each bin.}
\end{figure}

\begin{figure}[h]
\figfigincl{12 cm}{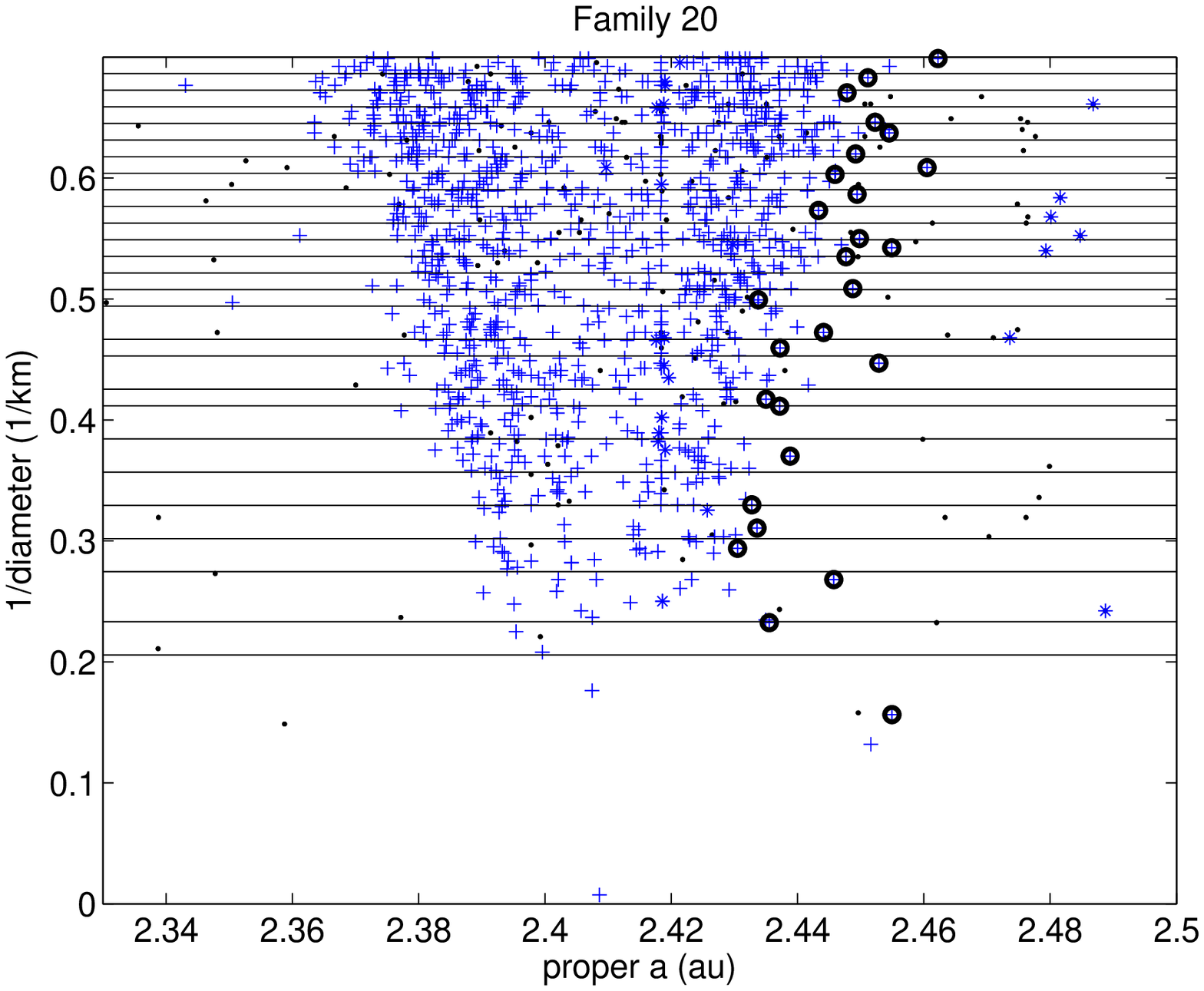}{Blow up of the bins for the
    outer side of the family of (20) Massalia. Circles are members of
    the family of (20) Massalia with the maximum value of proper $a$
    and the corresponding $1/D$ in each bin on the left side. Crosses,
    points and stars as in Fig.~\ref{fig:20_bins_in}.}
\end{figure}

\newpage
\subsection{Error Model and Weights}

The least squares fit, especially if it includes an outlier rejection
procedure, requires the existence of an error model for the values to
be fit. Until now there are no error models for the absolute magnitude
and the albedo, which are available for a large enough catalog of
asteroids. 

We have built a simple but realistic error model for $1/D$ computed
from the absolute magnitude $H$ (the formula is $D=1\,329\times
10^{-H/5}\, \times 1/\sqrt{p_v}$) by combining the effect of two terms
in the error budget: the error in the absolute magnitude with STD
$\sigma_{H}$ and the one in the geometric albedo with STD
$\sigma_{p_v}$.  The derivatives of $1/D$ with respect to these two
quantities are:
\[
\frac{\partial (1/D)}{\partial H} = 
\frac{log(10)}{5} \times \frac{1}{D} \ \ \; \ \ 
\frac{\partial (1/D)}{\partial p_v} = 
\left(\frac{1}{2 \times p_v}\right) \times \frac{1}{D} \ ,
\]
then the combined error has STD
\[
 \sigma_{1/D} = 
\sqrt{\left(\frac{\partial (1/D)}{\partial H}\;\sigma_{H}\right)^2 
+\left(\frac{\partial (1/D)}{\partial p_v}\;\sigma_{p_v}\right)^2}\ .
\]
To compute this error model we need to select three values: 1) the
common geometric albedo $p_v$ for all the family members, 2) the
dispersion with respect to this common albedo $\sigma_{p_v}$,  3)
the uncertainty in the absolute magnitude $\sigma_{H}$.

\begin{figure}[h!]
\figfigincl{12 cm}{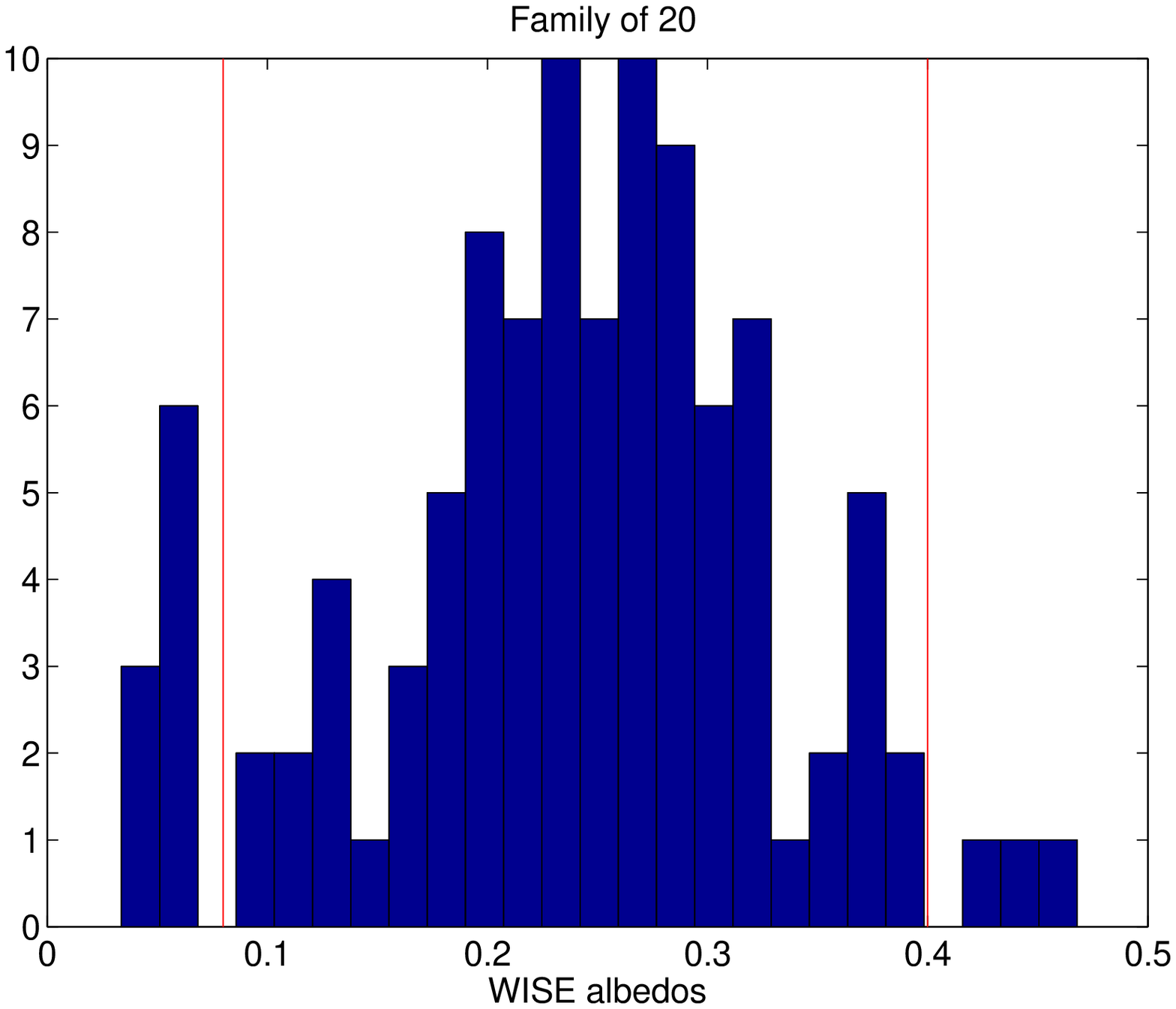}{Histogram of the ``significant'' WISE
  albedos for the dynamical family of (20) Massalia. The vertical
  lines show the values of the albedos used for the cut, leaving out
  values which should correspond to interlopers. In this and
    in many other cases the selection of the interlopers is simple:
    albedo $<0.1$ indicates C-complex asteroids and $>0.4$ values are
    likely to be affected by large errors.}
\end{figure}

For the first two, we select all the ``significant'' WISE albedos,
that is the values of the albedos greater than $3$ times their
standard deviations (with $S/N>3$). Then we cut the tails of this
distribution (see Figure \ref{fig:20_histalb}): $p_v$ is the mean and
$\sigma_{p_v}$ is the standard deviation of the values of the albedo
without the tails.
For the third value $\sigma_H$ we use the same for all the families
and the chosen value is $0.3$, see the discussion in
\citep{bigdata}[Sec. 2.2] and in \citep{pravecharris}.

The histograms such as Figure \ref{fig:20_histalb} are available for
all the families listed in Table~\ref{tab:albedo} at the Supplementary
material web site.

\begin{table}[h!]
\footnotesize
 \centering
 \caption{Family albedos: number and name of the family, albedo of the
   parent body with standard deviation and code of reference, maximum
   and minimum value for computing mean, mean and standard deviation
   of the albedo}
  \label{tab:albedo}
\medskip
  \begin{tabular}{|l|llc|llll|}
  \hline
number/          &albedo  &largest&   &     &family  &albedo &     \\
name             &value   &STD    &ref&min  & max &mean   &STD  \\
\hline
158 Koronis      &  0.144 &  0.009 &W& 0.07& 0.50& 0.240 & 0.061\\
24  Themis       &  0.064 &  0.016 &W&     & 0.12& 0.069 & 0.019\\
847 Agnia        &  0.147 &  0.012 &W& 0.10& 0.40& 0.242 & 0.056\\
1726 Hoffmeister &  0.036 &  0.007 &W&     & 0.10& 0.048 & 0.013\\
668 Dora         &  0.073 &  0.009 &W&	   & 0.10& 0.058 & 0.014\\
434 Hungaria     &  0.380 &        &S&	   &     & 0.380 & 0.100\\
480 Hansa        &  0.249 &  0.024 &I& 0.10& 0.45& 0.286 & 0.068\\
808 Merxia       &  0.165 &  0.021 &W& 0.10& 0.40& 0.248 & 0.055\\
3330 Gantrisch   &  0.048 &  0.010 &W&     &     & 0.047 & 0.012\\
10955 Harig      &        &        & &     &     & 0.251 & 0.068\\
293 Brasilia     &  0.033 &  0.007 &W& 0.10& 0.27& 0.174 & 0.042\\
569 Misa         &  0.030 &  0.001 &I&     & 0.10& 0.058 & 0.016\\
1128 Astrid      &  0.046 &  0.018 &W&     &     & 0.052 & 0.014\\
845 Naema        &  0.072 &  0.019 &W&     & 0.10& 0.065 & 0.014\\
\hline
4 Vesta          &  0.423 &  0.053 &I& 0.15& 0.60& 0.355 & 0.099\\
15 Eunomia       &  0.206 &  0.055 &W&     & 0.50& 0.260 & 0.083\\
20 Massalia      &  0.210 &  0.030 &I& 0.08& 0.40& 0.249 & 0.070\\
10 Hygiea        &  0.058 &  0.005 &W& 0.02& 0.15& 0.073 & 0.022\\
31 Euphrosyne    &  0.045 &  0.045 &W&     & 0.10& 0.061 & 0.015\\
3  Juno          &  0.238 &  0.025 &I& 0.10& 0.40& 0.253 & 0.055\\
163 Erigone      &  0.033 &  0.004 &W&     & 0.10& 0.055 & 0.013\\
\hline
3815 K\"onig     &  0.056 &  0.004 &W&     & 0.15& 0.051 & 0.014\\
396 Aeolia       &  0.139 &  0.025 &W&     &     & 0.106 & 0.028\\
606 Brangane     &  0.089 &  0.012 &W&     &     & 0.121 & 0.028\\
1547 Nele        &  0.313 &  0.040 &A& 0.15&     & 0.355 & 0.064\\
18405 1993FY$_{12}$&       &        & & 0.10&     & 0.184 & 0.042\\ 
\hline
170 Maria        &  0.160 &  0.007 &I&     &     & 0.261 & 0.084\\
93 Minerva       &  0.073 &  0.004 &I& 0.10& 0.50& 0.277 & 0.096\\
2076 Levin       &  0.557 &  0.318 &W& 0.10& 0.40& 0.202 & 0.070\\
3827 Zdenekhorsky&  0.104 &  0.008 &W&     & 0.12& 0.074 & 0.020\\
1658 Innes       &  0.224 &  0.037 &W& 0.10& 0.43& 0.264 & 0.064\\
375 Ursula       &  0.049 &  0.001 &A&     & 0.10& 0.062 & 0.015\\
\hline
\end{tabular}
\end{table}

In Table~\ref{tab:albedo} we show the albedo value of the namesake
asteroid, with its uncertainty and the appropriate reference: W for
WISE data~\citep{Masiero_WISE}, I for IRAS, S for
\citep{shepard08}, and A for AKARI. In some cases albedo data are not
available.  The columns $5$ and $6$ contain the value of the albedo
used for the cut of the tails, and the last two columns are the mean
albedo and the standard deviation.

Two discordant results from the albedo analysis of the dynamical
families are easily appreciated from Table~\ref{tab:albedo}. (93)
Minerva and (293) Brasilia are interlopers in the dynamical families
for which they are namesake, as shown by albedo data outside of the
family range. Indeed, in the following of this paper we are going to
speak of the family 1272 (Gefion) instead of 93, and of the family
1521 (Seinajoki) instead of 293; both are obtained by removing
interlopers selected because of albedo data, and the namesake is the
lowest numbered after removing the interlopers.

For many families we have proceeded in the same way, that is removing
interlopers clearly indicated by an albedo discordance. The list of
these interlopers for each family is in the Supplementary material.

In some cases we have joined two dynamical families for the purpose of
mean albedo computation: 2076 includes 298, 163 include 5026, 10955
includes 19466\footnote{However, there is only 1 significant WISE
  albedo among members of family 19466.}.  Family 847 includes the
subfamily 3395: the same mean albedo was used for both, although (847)
has albedo $0.147\pm 0.01$ and (3395) $0.313\pm 0.05$, which are on
the opposite side of the mean. Also 569 in Table~\ref{tab:albedo}
includes the subfamily 15124.

The family of (434) Hungaria is a difficult case: some WISE data exist
for its family members, but they are of especially poor quality. Thus
we have used for all the albedo derived from radar data
\citep{shepard08}, and assumed a quite large dispersion ($0.1$).
     
\subsection{Outlier Rejection and Quality Control}
\label{s:outlier_rejection}

The algorithm for differential corrections used for the computation of
the slopes includes an automatic outlier rejection scheme, as in
~\citep{carpino03}. Both the use of an explicit error model for the
observations and the fully automatic outlier rejection procedure are
implemented in the free software OrbFit\footnote{Distributed at
  http://adams.dm.unipi.it/orbfit/} and are used for the orbit
determination of the asteroids included in the NEODyS and AstDyS
information systems\footnote{http://newton.dm.unipi.it/neodys/ and
  http://hamilton.dm.unipi.it/astdys/ respectively.}. Thus, although
the application of these methods to the computation of family ages is
new, this is a very well established procedure on which we have a lot
of experience.

In practice, outlier rejection is performed in an iterative way. At
each iteration, the program computes the residuals of all the
observations, their expected covariance and the corresponding $\chi^2$
value. If we can assume that the observation errors have a normal
distribution, to mark an observation as an outlier we can compare the
$\chi^2$ value of the post-fit residual with a threshold value
$\chi^2_{rej}$: the observation is discarded if
$\chi_{i}^2>\chi^2_{rej}$. At each iteration it is also necessary to
check if a given observation, that we have previously marked as an
outlier, should be recovered. Therefore, the program selects an
outlier to be recovered if for the non-fitted residual
$\chi_i^2<\chi_{rec}^2$. The current values for $\chi^2_{rej}$ and
$\chi_{rec}^2$ are $10$ and $9$, respectively.

During each iteration of the linear regression we compute the
residuals, the outliers, the RMS of the weighted residuals and the
Kurtosis of the same weighted residuals. Our method converges if there
is an iteration without additional outliers. All these data are
reported in Table 1 of the Supplementary material. Besides the
automatic outlier rejections, some \textit{interlopers} have been
manually removed when there was a specific evidence that they do not
belong to the collisional family, e.g., based upon WISE data: also these
manual rejections are detailed in the Supplementary material.

\section{Results}
\label{s:results}

\subsection{Fragmentation Families}

The results of the fit for the slopes of the V-shape are described in
Table~\ref{tab:slope_frag} for the families of the fragmentation
type. To define fragmentation families, we have used the (admittedly
conventional) definition that the volume of the family without the
largest member has to be more than $12\%$ of the total. This
computation has been done after removing the interlopers (by physical
properties) and the outliers (removed in the fit), and is based on $D$
computed with the mean albedo $p_v$. Comments for some of the cases
are given below.

\begin{table}[h!]
\footnotesize
 \centering
 \caption{Slopes of the V-shape for the fragmentation families: family
   number/name, number of dynamical family members, side, slope ($S$),
   inverse slope ($1/S$), standard deviation of the inverse slope,
   ratio OUT/IN of $1/S$, and standard deviation of the ratio.}
  \label{tab:slope_frag}
\medskip
  \begin{tabular}{|lr|crrl|ll|}
  \hline
number/ & no.      & side    & S       & $1/S$  & STD   & ratio& STD \\
name    & members  &         &         &        &$1/S$  &      & ratio\\
\hline
158 Koronis       &6130 & IN & -1.647  & -0.608 & 0.089 &    &    \\ 
                  &     & OUT&  1.755  &  0.570 & 0.069 &0.94&0.18\\
24  Themis        &4329 & IN & -0.720  & -1.390 & 0.385 &    &    \\
                  &     & OUT&  0.477  &  2.096 & 0.326 &1.51&0.48\\
847 Agnia         &2395 & IN & -2.882  & -0.347 & 0.072 &    &    \\
                  &     & OUT&  4.381  &  0.228 & 0.034 &0.66&0.17\\
3395 Jitka        &     & IN &-21.387  & -0.047 & 0.009 &    &    \\
                  &     & OUT& 21.214  &  0.047 & 0.008 &1.01&0.26\\
1726 Hoffmeister  &1560 & IN & -5.026  & -0.199 & 0.028 &    &    \\
                  &     & OUT&  5.212  &  0.192 & 0.025 &0.96&0.18\\
668 Dora          &1233 & IN & -3.075  & -0.325 & 0.053 &    &    \\
                  &     & OUT&  3.493  &  0.286 & 0.086 &0.88&0.30\\
434 Hungaria      &1187 & IN &-14.855  & -0.067 & 0.006 &    &    \\
                  &     & OUT& 15.293  &  0.065 & 0.003 &0.97&0.10\\
480 Hansa         &960  & IN & -3.710  & -0.270 & 0.109 &    &    \\
                  &     & OUT&  3.064  &  0.326 & 0.040 &1.21&0.51\\
808 Merxia        &924  & IN & -8.400  & -0.119 & 0.010 &    &    \\
                  &     & OUT&  8.963  &  0.112 & 0.008 &0.94&0.10\\
3330 Gantrisch    &723  & IN & -3.986  & -0.251 & 0.061 &    &    \\
                  &     & OUT&  3.552  &  0.282 & 0.075 &1.12&0.40\\
10955 Harig \&    &517  & IN & -6.515  & -0.154 & 0.027 &    &    \\
19466 Darcydiegel &153  & OUT&  5.853  &  0.171 & 0.050 &1.11&0.38\\
1521 Seinajoki    &545  & IN & -8.454  & -0.118 & 0.023 &    &    \\
                  &     & OUT& 23.507  &  0.043 & 0.005 &0.36&0.08\\
569 Misa          &441  & IN & -5.0376 & -0.199 & 0.151 &    &    \\
                  &     & OUT&  6.5380 &  0.153 & 0.052 &0.77&0.64\\
15124 2000EZ$_{39}$&     & IN &-14.422  & -0.069 & 0.006 &    &    \\
                  &     & OUT& 14.337  &  0.070 & 0.007 &1.01&0.14\\
1128 Astrid       &436  & IN &-11.339  & -0.088 & 0.006 &    &    \\
                  &     & OUT& 11.434  &  0.088 & 0.007 &0.99&0.10\\
845 Naema         &286  & IN &-11.715  & -0.085 & 0.011 &    &    \\
                  &     & OUT& 10.741  &  0.093 & 0.004 &1.09&0.15\\
\hline
\end{tabular}
\end{table}

\begin{itemize}

\item For family 158 (Koronis) the values of the inverse
  slope $1/S$ on the two sides are consistent, that is the ratio is
  within a standard deviation from $1$: this indicates that we are
  measuring the age of a single event. The well known subfamily of
  (832) Karin, with a recent age, does not affect the slopes.

\item Family 24 (Themis) has the well known subfamily of (656) Beagle
  near the center of the V-shape, thus it does not affect the
  slopes. The values IN and OUT are not the same but the difference
  has very low statistical significance.  The low accuracy of the IN
  slope determination is due to the fact that the 11/5 resonance cuts
  the V-shape too close to the center, sharply reducing the useful
  range in $D$.

\item For family 847 (Agnia) we have estimated also the slopes for the
  subfamily 3395.  847 has discordant slope values on the two sides, but the
  OUT one has too few points, being affected by 3395. Thus we consider
  as the true value the one obtained on the IN side. 3395 has a very
  good fit but with many outliers, which can be explained as members
  of 847 but not 3395. Anyway the inverse slopes are significantly
  lower for 3395; since the ages are proportional to the inverse
  slopes, this indicates an age younger by a factor $7.42\pm 2.06$
  (based upon the IN values).

\item Family 1726 (Hoffmeister) has an especially complicated dynamics
  on the IN side, due to both the nonlinear secular resonance
  $g+s-g_6-s_6$ and the proximity with (1) Ceres, see the discussion
  in \citep{bigdata}[Sec. 4.1]. However, the results on the two slopes
  are perfectly consistent: this is in agreement with what was claimed
  by \cite{delisle_laskar}, namely that the Yarkovsky effect prevails
  over the chaotic effects induced by close approaches (also by the 1-1
  resonance) with Ceres, in the range of sizes which is relevant for the fit.

\item For the family 480 (Hansa) the slope for the IN side has lower
  quality, probably due to 3/1 resonance. It is a marginal
  fragmentation with $14\%$ of the total volume, excluding (480) Hansa
  itself.

\item Family 808 (Merxia) is a fragmentation with a dominant largest
  member ($64\%$ in volume), thus (808) must not be
  included in the fit.

\item For the family 3330 (Gantrisch) it has been difficult to
  compute a slope for the IN side, because of the irregular shape of
  the low $a$ border resulting in few data to be fit. 

\item Family 10955 (Harig) can be joined with family 19466: in this
  way two one-sided families form a single V-shape: this join is
  confirmed by the two slopes being consistent. Thus one collisional
  family is obtained from two dynamical families. This merge was
  already suggested in \citep{bigdata}[Sec. 4.3.2], based on the family
  box overlap (by $40\%$).

\item Family 1521 (Seinajoki) appears to have two discordant
  slopes: in the projection $(a, \sin{I})$ a bimodality appears in
  the family shape. We draw from this the conclusion that there are
  two collisional families, the one on the IN side being older.

\item The family 569 (Misa) is a marginal fragmentation (fragments
  account for $19\%$ of the total volume). The ratio of the IN and OUT
  slopes is not significantly different from 1, mostly because of the
  low accuracy of the IN value. (15124) 2000 EZ$_{39}$ appears to be
  the largest fragment of a fragmentation subfamily inside the family
  569: the inverse slopes are significantly lower, indicating an age
  younger by a factor $2.19\pm 0.78$ with respect to 569 (based upon
  the OUT values).



\end{itemize}

\subsection{Cratering Families}

The results of the fit for the slopes of the V-shape are described in
Table~\ref{tab:slope_crater} for the families of the cratering type,
defined by a volume of the family without the largest member $<12\%$
of the total.  Comments for some of the cases are given below.

\begin{table}[h!]
\footnotesize
 \centering
 \caption{Slope of the V-shape for the cratering families. Columns as
   in Table~\ref{tab:slope_frag}.}
  \label{tab:slope_crater}
\medskip
  \begin{tabular}{|lr|crrl|ll|}
  \hline
number/ & no.   &side & S & $1/S$ & STD   & ratio& STD\\
name    &members&      &   &       & $1/S$ &      & ratio\\
\hline
4 Vesta        &8620 & IN & -2.983  & -0.335  & 0.040 &&\\
               &     & OUT&  1.504  &  0.665  & 0.187 &1.98&0.61\\
15 Eunomia     &7476 & IN & -1.398  & -0.715  & 0.057 &&\\
               &     & OUT&  2.464  &  0.406  & 0.020 &0.57&0.05\\
20 Massalia    &5510 & IN &-15.062  & -0.066  & 0.003 &&\\
               &     & OUT& 14.162  &  0.071  & 0.006 &1.06&0.10\\
10 Hygiea      &2615 & IN & -1.327  & -0.754  & 0.079 &&\\
               &     & OUT&  1.329  &  0.752  & 0.101 &1.00&0.17\\
31 Euphrosyne  &1137 & IN & -1.338  & -0.747  & 0.096 &&\\
               &     & OUT&  1.507  &  0.663  & 0.081 &0.89&0.16\\
3 Juno         &960  & IN & -5.261  & -0.190  & 0.038 &&\\
               &     & OUT&  7.931  &  0.126  & 0.049 &0.66&0.29\\
163 Erigone \& &429  & IN & -7.045  & -0.142  & 0.035 &&\\
5026 Martes    &380  & OUT&  6.553  &  0.153  & 0.013 &1.08&0.28\\
\hline
\end{tabular}
\end{table}

\begin{itemize}

\item Family 4 (Vesta) has two discordant slopes on the IN and OUT
  sides. As already suggested in \citep{bigdata}[Sec. 7.2], this should
  be interpreted as the effect of two distinct collisional families,
  with significantly different ages. The estimated ratio of the slopes
  provides a significant estimate of the ratio of the ages, because
  the Yarkovsky calibration is common to the two subfamilies,
  corresponding to two craters on Vesta. 

\item Family 15 (Eunomia) has a subfamily which determines the OUT
  slope, the ratio of the slopes gives a good estimate of the ratio of
  the ages, because of the common calibration. The interpretation as
  two collisional families, proposed in \citep{bigdata}[Sec. 7.4], is
  thus confirmed.

\item Family 10 (Hygiea) has a shape (especially in the proper $(a,e)$
  projection) from which we could suspect two collisional events, but
  the IN and OUT slopes not just consistent but very close suggest a
  single collision.
  
\begin{figure}[h!]
\figfigincl{10 cm}{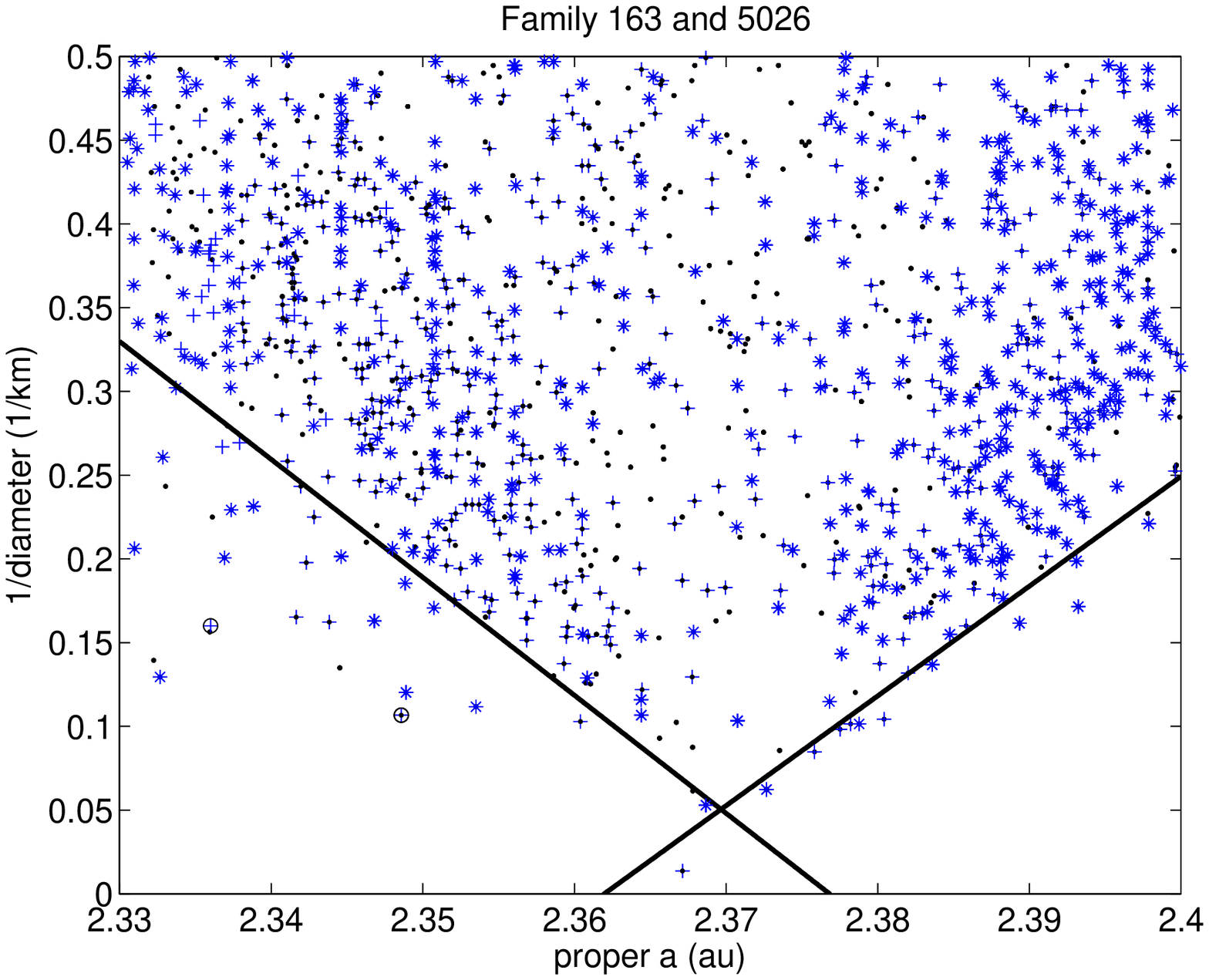}{V-shape fit for the join of
  families of (163) Erigone and (5026) Martes. The IN slope is fit to
  members of 163, the OUT slope to members of 5026, but the two values
  are consistent. The central depleted region explains why the two
  families have no intersection: they are \textit{joined} but not
  \textit{merged}.}
\end{figure}



\item For family 3 (Juno) the IN and OUT slopes are discordant, but due
  to the low relative accuracy of the slopes the difference is
  marginally significant.  The number density as a function of proper
  $a$ is asymmetric, more dense on the OUT side.

\item Family 163 (Erigone) can be joined with 5026 (Martes), with
  (163) as parent body for both (marginally within the cratering
  definition, fragments forming $11\%$ of the total volume). This is
  confirmed by similar albedo (dark in a region dominated by brighter
  asteroids) and by very consistent slopes of the IN side (formed by
  family 163) and of the OUT side (formed by 5026), see
  Figure~\ref{fig:163_5026_vshapea_paper}. There is a very prominent
  gap in the center, which explains why we have found no
  intersections; it should be due to the YORP effect; see
  Section~\ref{s:open}. Again one collisional family is obtained from
  two dynamical families.

\end{itemize}

\subsection{Young Families}

We define as \textit{young families} those with an estimated age of
$<100$ My; thus the inverse slopes are much lower than those of the
previous tables. These can be both fragmentations and craterings. The
results of the fit are described in Table~\ref{tab:slope_young}.

\begin{table}[h!]
\footnotesize
 \centering
 \caption{Slope of the V-shape for the young families. Columns as in
   Table~\ref{tab:slope_frag}.}
  \label{tab:slope_young}
\medskip
  \begin{tabular}{|lr|crrl|ll|}
  \hline
number/ &no.    & side& S & $1/S$ & STD    & ratio & STD\\
name    &members&     &   &       &  $1/S$ &       & ratio\\
\hline
3815 K\"onig    &340 & IN & -31.892  & -0.031  & 0.004 &&\\
                &    & OUT&  31.364  &  0.032  & 0.004 &1.02&0.17\\
396 Aeolia      &306 & IN & -32.358  & -0.031  & 0.005 &&\\
                &    & OUT&  35.556  &  0.028  & 0.005 &0.91&0.22\\
606 Brangane    &192 & IN & -54.027  & -0.019  & 0.002 &&\\
                &    & OUT&  60.374  &  0.017  & 0.003 &0.89&0.17\\
1547 Nele       &152 & IN &-201.336  & -0.005  & 0.0008 &   &\\
                &    & OUT& 187.826  &  0.005  & 0.002 & 1.07 & 0.44\\   
18405 1993FY$_{12}$& 102& IN& -34.189  & -0.029  & 0.01  &      &     \\
                &    & OUT&  34.456  &  0.029  & 0.005 & 0.99 & 0.36\\
\hline
\end{tabular}
\end{table}

These families have a comparatively low number of members, but because
they also have a small range of proper $a$ values a significant slope
fit is possible. In particular we have introduced the three last
families in the Table with $<250$ members.  

Few comments: families 396 (Aeolia) and 606 (Brangane) are craterings,
all the others are fragmentations. The family 1547 (Nele) is a
marginal fragmentation, with $19\%$ of volume outside (1547); it is
known to be very young \citep{nesv03}, it has been included to test
the applicability of the V-shape method to recent families (see
Sec.~\ref{s:ages_young}).

\subsection{One Side}
\label{s:oneside}

The one-sided families are those for which we cannot identify one of
the two sides of the V-shape. The results of the fit are described in
Table~\ref{tab:slope_one}.

The families of this type can be due to fragmentations and
craterings: in most cases there is no dominant largest fragment, and
they might have had parent bodies disappeared in the resonance which
also wiped out one of the sides, thus we do not really know.

\begin{table}[h!]
\footnotesize
 \centering
 \caption{Slopes of the V-shape for the one-sided families: family
   number/name, number of dynamical family members, side, slope ($S$),
   inverse slope ($1/S$), standard deviation of the inverse slope.}
  \label{tab:slope_one}
\medskip
  \begin{tabular}{|lr|crrl|}
  \hline
number/ &no.    &side & S & $1/S$ & STD \\
name    &members&     &   &       & $1/S$\\
\hline
170 Maria        &1431 & OUT &  1.487  &  0.672  & 0.059\\
1272 Gefion      &1341 & IN  & -2.594  & -0.386  & 0.094\\ 
2076 Levin       &1237 & OUT &  7.080  &  0.141  & 0.023\\
3827 Zdenekhorsky&794  & IN  &-10.871  & -0.092  & 0.009\\
1658 Innes       &606  & OUT &  6.006  &  0.167  & 0.011\\
375 Ursula       &335  & IN  & -0.516  & -1.938  & 0.426\\
\hline
\end{tabular}
\end{table}

\begin{itemize}

\item The family 170 (Maria) has a possible subfamily for low proper
  $a$ (no effect on the OUT slope). There is no dominant largest
  fragment, thus it could be either a fragmentation or a
    cratering, in the latter case with parent body removed by the 3/1
  resonance.

\item For the family 1272 (Gefion) there is no dominant largest
  fragment, thus the same argument applies, with possible parent body
  removal by the 5/2 resonance.

\item For family 2076 (Levin) the possibility of merging with families
  298 (Baptistina) and 883 has been discussed in
  \citep{bigdata}[Sec. 4.1]. Joining Baptistina does not change the
  slopes; joining 883 would result in a two-sided V-shape, with a gap
  due to the 7/2 resonance in between; however, the two slopes would
  be very different. All three dynamical families (for which we
  already have some intersections) could be considered as a single
  complex dynamical family, but still they would belong to different
  collisional families with different ages. The slope (thus the age)
  we have computed belongs to the event generating only the 2076
  family.  There are not enough significant physical data on the
  members of these families\footnote{E.g., (2076) has WISE data
    $p_v=0.56\pm 0.32$.}, not even on the comparatively large (298),
  to help us in disentangling this complex case.

\item Family 1658 (Innes) is the largest fragment but it is not
  dominant in size, thus we cannot distinguish between fragmentation
  and cratering with parent body removed by the 3/1 resonance.
 
\item (375) Ursula is an outlier in the fit for the IN slope of
  375. This can have two interpretations. Either (375) is the largest
  fragment of a marginal fragmentation (fragments are $23\%$ of total
  volume), in which case it is correct not to include it in the slope
  fit, or (375) is an interloper and the family could have had a
  parent body later disappeared in the 2/1 resonance.  Unfortunately,
  it is difficult to use albedo data to help on this, because there is
  no albedo contrast with the background.

\end{itemize}

\section{Age Estimation}
\label{s:ages_est}

\subsection{Yarkovsky Calibrations}
\label{s:yarko_cal}

The method we use to convert the inverse slopes from the V-shape fit
into family ages has been established in \citep{bigdata}[Sec. 5.2],
and consists in finding a \textit{Yarkovsky calibration}, which is the
value of the Yarkovsky driven secular drift $da/dt$ for an
hypothetical family member of size $D=1$ km and with spin axis
obliquity (with respect to the normal to the orbital plane) $0^\circ$
for the OUT side and $180^\circ$ for the IN side. Since the inverse
slope is the change $\Delta(a)$ accumulated over the family age by a
family member with unit $1/D$, the age is just $\Delta(t)=\Delta(a)/(da/dt)$.

The question is how to produce the Yarkovsky calibration. As discussed
in \citep{bigdata}[Sec. 5.2.6], this can be done in different ways
depending upon which data are available. Unfortunately for main belt
asteroids there are too few data to compute any calibration: indeed, a
measured $da/dt$ is available for not even one main belt object. The
solution we have used was to extrapolate from the data available for
Near Earth Asteroids. The best estimate available for $da/dt$ is the
one of asteroid (101955) Bennu, with a $S/N\simeq 200$ \citep{bennu}.
By suitable modeling of the Yarkovsky effect, by using the available
thermal properties measurements, the density of Bennu has been
estimated as $\rho_{Bennu}=1.26\pm 0.07$ g/cm$^3$.  Bennu is a B-type
asteroid, thus it is possible to compute its porosity by comparison
with the very large asteroid (704) Interamnia which is of the same
taxonomic type and has a reasonably well determined bulk density
\citep{carry12}.

\begin{table}[h!]
\footnotesize
 \centering
 \caption{Benchmark asteroids for the density of a taxonomic
   type: number/name, taxonomic type, densities as in~\citep{carry12}
   with their uncertainties, densities at $1$ $km$.}
  \label{tab:benchmark}
\medskip
  \begin{tabular}{|l|ccc|c|}
  \hline
number/           & tax  & $\rho$& $STD(\rho)$ & $\rho$  \\
name              & type &       &             &($1$ $km$)\\
                  &      &       &             &\\
\hline
4   Vesta         &  V   &  3.58 &   0.15      &  2.30 \\
10  Hygiea        &  C   &  2.19 &   0.42      &  1.41 \\
15  Eunomia       &  S   &  3.54 &   0.20      &  2.275\\ 
216 Kleopatra     &  Xe  &  4.27 &   0.15      &  2.75 \\
704 Interamnia    &  B   &  1.96 &   0.28      &  1.26 \\  
\hline
\end{tabular}
\end{table}

In Table~\ref{tab:benchmark} we list the data on benchmark large
asteroids with known taxonomy and density. For the other
  taxonomic classes we estimate the density at $D=1$ km by assuming
  the same porosity of Bennu and the same composition as the largest
  asteroid of the same taxonomic class. Thus in the Table the density
at $D=1$ km for B class is the one of Bennu from \citep{bennu}, the
ones for the other classes are obtained by scaling.

Once an estimate of the density $\rho$ is available, the scaling
formula can be written as:
\[
\frac{da}{dt} = \left.\frac{da}{dt}\right|_{Bennu}
\frac{\sqrt{a}_{(Bennu)}(1-e^2_{Bennu})}{\sqrt{a}(1-e^2)}
\frac{D_{Bennu}}{D}\frac{\rho_{Bennu}}{\rho}
\frac{\cos(\phi)}{\cos(\phi_{Bennu})}\frac{1-A}{1-A_{Bennu}}
\]
where $D=1$ km used in this scaling formula is not the diameter of an
actual asteroid, but it is the reference value corresponding to the
inverse slope; we also assume $\cos(\phi)=\pm 1$, depending upon the
IN/OUT side.

The additional terms which we would like to have in the scaling
formula are thermal properties, such as thermal inertia or thermal
conductivity: the problem is that these data are not available. To
replace the missing thermal parameters with another scaling law would
not give a reliable result, also because of the strong nonlinearity of
the Yarkovsky effect as a function of the conductivity, as shown in
\citep{vok_2000}[Figure 1].

\begin{table}[p]
\footnotesize
 \centering
 \caption{Data for the Yarkovsky calibration: family number and name,
   proper semimajor axis $a$ and eccentricity $e$ for the inner and
   the outer side, 1-A, density value $\rho$ at $1$ km, taxonomic
   type, a flag with values m (measured) a (assumed) g (guessed), and
   the relative standard deviation of the calibration.}
  \label{tab:calibration}
\medskip
  \begin{tabular}{|l|llll|cl|ccc|}
  \hline
number/          & proper& proper& proper& proper& 1-A & $\rho$    & tax. & flag&rel.\\
name             & $a$   & $e$   & $a$   & $e$   &     & ($1$ $km$)& type & m   &STD \\
                 & IN    & IN    & OUT   & OUT   &     &           &      &     & \\  
\hline
158 Koronis      & 2.83  & 0.044 & 2.93  & 0.06  & 0.92&  2.275    &  S   & m & 0.20\\
24  Themis       & 3.085 & 0.14  & 3.23  & 0.135 & 0.98&  1.41     &  C   & m & 0.20\\ 
847 Agnia        & 2.73  & 0.07  & 2.81  & 0.07  & 0.92&  2.275    &  S   & m & 0.20\\
3395 Jitka       & 2.762 & 0.07  & 2.81  & 0.07  & 0.92&  2.275    &  S   & m & 0.20\\  
1726 Hoffmeister & 2.76  & 0.05  & 2.8   & 0.046 & 0.98&  1.41     &  C   & a & 0.25\\
668 Dora         & 2.76  & 0.19  & 2.8   & 0.197 & 0.98&  1.41     &  C   & a & 0.25\\
434 Hungaria     & 1.92  & 0.07  & 1.97  & 0.07  & 0.87&  2.75     &  Xe  & g & 0.30\\
480 Hansa        & 2.55  & 0.04  & 2.69  & 0.035 & 0.91&  2.45     &  S   & m & 0.20\\
808 Merxia       & 2.71  & 0.135 & 2.78  & 0.13  & 0.92&  2.45     &  S   & m & 0.20\\
3330 Gantrisch   & 3.13  & 0.195 & 3.16  & 0.198 & 0.98&  1.41     &  C   & g & 0.30\\
10955 Harig \&   & 2.67	 & 0.016 &       &       & 0.92&  2.275    &  S   & g & 0.30\\
19466 Darcydiegel&       &       & 2.77  & 0.009 & 0.92&  2.275    &  S   & g & 0.30\\
1521 Seinajoki   & 2.84  & 0.12  & 2.866 & 0.123 & 0.94&  2.275    &  S   & g & 0.30\\
569 Misa         & 2.63  & 0.177 & 2.69  & 0.175 & 0.98&  1.41     &  C   & a & 0.25\\
15124 2000EZ$_{39}$& 2.64  & 0.178 & 2.67  & 0.177 & 0.98&  1.41     &  C   & a & 0.25\\
1128 Astrid      & 2.767 & 0.048 & 2.81  & 0.048 & 0.98&  1.41     &  C   & m & 0.20\\
845 Naema        & 2.92  & 0.035 & 2.95  & 0.036 & 0.98&  1.41     &  C   & m & 0.20\\
\hline
4 Vesta          & 2.27  & 0.09  & 2.44  & 0.11  & 0.88&  2.3      &  V   & m & 0.20\\
15 Eunomia       & 2.53  & 0.15  & 2.69  & 0.15  & 0.92&  2.275    &  S   & m & 0.20\\
20 Massalia      & 2.35  & 0.162 & 2.46  & 0.162 & 0.92&  2.275    &  S   & m & 0.20\\
10 Hygiea        & 3.08  & 0.13  & 3.24  & 0.11  & 0.98&  1.41     &  C   & m & 0.20\\
31 Euphrosyne    & 3.11  & 0.17  & 3.2   & 0.21  & 0.98&  1.41     &  C   & m & 0.20\\
3 Juno           & 2.62  & 0.235 & 2.69  & 0.235 & 0.92&  2.275    &  S   & m & 0.20\\
163 Erigone \&   & 2.34  & 0.208 &       &       & 0.98&  1.41     &  C   & m & 0.20\\
5026 Martes      &       &       & 2.37  & 0.207 & 0.98&  1.41     &  C   & m & 0.20\\
\hline
3815 K\"onig     & 2.57  & 0.13  & 2.58  & 0.14  & 0.98& 1.41      &  C   & a & 0.25\\
396 Aeolia       & 2.735 & 0.168 & 2.743 & 0.167 & 0.97& 2.75      &  Xe  & a & 0.25\\
606 Brangane     & 2.579 & 0.18  & 2.59  & 0.18  & 0.96& 2.275     &  S   & m & 0.20\\
1547 Nele        & 2.64  & 0.269 & 2.646 & 0.269 & 0.88& 2.275     & S    & g & 0.30\\  
18405 1993FY$_{12}$& 2.83 & 0.106 & 2.86  & 0.106 & 0.94& 2.275     & S    & g & 0.30\\
\hline
170 Maria        &       &       & 2.65  & 0.08  & 0.91& 2.275     &  S   & m & 0.20\\
1272 Gefion      & 2.74  & 0.13  &       &       & 0.92& 2.275     &  S   & a & 0.25\\
2076 Levin       &       &       & 2.31  & 0.14  & 0.93& 2.275     &  S   & g & 0.30\\
3827 Zdenekhorsky& 2.71  & 0.087 &       &       & 0.98& 1.41      &  C   & m & 0.20\\
1658 Innes       &       &       & 2.61  & 0.17  & 0.91& 2.275     &  S   & g & 0.30\\
375 Ursula       & 3.13  & 0.08  &       &       & 0.98& 1.41      &  C   & m & 0.20\\
\hline
\end{tabular}
\end{table}

We are not claiming this is the best possible calibration for each
family. However, for generating a homogeneous set of family ages, we
have to use a uniform method for all. To improve the calibration (thus
to decrease the uncertainty of the age estimate) for a specific family
is certainly possible, but requires a dedicated effort in both
acquiring observational data and modeling. E.g., the Yarkovsky effect
could be measured from the orbit determination for a family member
(going to be possible with data from the astrometric
mission GAIA), thermal properties could be directly measured with
powerful infrared telescopes, densities can be
derived for binaries by using radar observations, for the cases with
missing taxonomy it could be obtained by
spectrometry/infrared/polarimetry. A good example is given by a very
recent event: on January 26, 2015 the asteroid (357439) 2004 BL$_{86}$
had a very close approach to the Earth, with minimum distance $0.008$
au. Thus it has been possible by radar to confirm that it has a
satellite, and to measure its diameter; infrared observations allowed
to assign this asteroid to the taxonomic class V. When all the data
are analyzed and published, we expect to have for (357439) an
estimated density (from the satellite orbit and the volume, both from
the radar data). This could provide a Yarkovsky calibration,
specifically for the Vesta families, significantly better than the one
of this paper.

This implies that the main results of this paper are the inverse
slopes, from which the ages can continue to be improved as better
calibration data become available.

In Table~\ref{tab:calibration} we are summarizing the data used to
compute the calibration. The eccentricity used in the calibration is
selected, separately for the IN and OUT side, as an approximate
average of the values of proper eccentricity for the family members
with proper semimajor axis close to the limit. It is clear that the
extrapolation from Near Earth to main belt asteroids introduces a
model uncertainty, which is not the same in all cases. If a family has
a well determined taxonomic type, which corresponds to one of the
benchmark asteroids, our computation of the calibration is based on
actual data and we assign to this case a comparatively low relative
calibration STD of $0.2$; these cases are labeled with the code
``m''. We have also estimated the Bond albedo A, which is used in the
scaling, from the mean geometric albedo $p_v$ by WISE. For subfamilies
3395 (inside 847) and 15124 (inside 569) we have assumed the same
taxonomy as the larger family.

Then there are cases in which the taxonomic class is similar, but not
identical to the one of the benchmark. (1726) is of type Cb, (668) of
type Ch in the SMASSII classification, both assimilated to a generic C
type; (808) is Sq, (1272) is SI in SMASSII, (1658) is AS in the Tholen
classification, all assimilated to a generic S type. These are labeled
with the code ``a'' and we have assigned a relative STD of $0.25$.

Finally we have 7 cases in which we do not have taxonomic data at all,
but just used the mean WISE albedo of Table~\ref{tab:albedo} to guess
a simplistic classification into a C vs. S complex. These are labeled
``g'' and have a relative STD of $0.3$. Thus these are the worst cases
from the point of view of age uncertainty, but they are the easiest to
improve by observations.

\subsection{Ages and their Uncertainties}
\label{s:ages}

The results on the ages are presented in
Tables~\ref{tab:age_frag}--\ref{tab:age_one}, each containing the
Yarkovsky calibration, computed with the data of
Table~\ref{tab:calibration}, the estimated age and three measures of the
age uncertainty. 

The first uncertainty is the standard deviation of the inverse slope,
as output from the least square fit, divided by the calibration. The
second is the age uncertainty due to the calibration uncertainty from
the last column of Table~\ref{tab:calibration}: this relative
uncertainty is multiplied by the estimated age. The third is the
standard deviation of the age, obtained by combining quadratically the
STD from the fit with the STD from the calibration.

The first uncertainty is useful when comparing ages which can use the
same calibration, such as ages from the IN and from the OUT side (as
shown in the last two columns of
Tables~\ref{tab:slope_frag}--\ref{tab:slope_one}); this can be applied
also to the cases of subfamilies. The third uncertainty is applicable
whenever the absolute age has to be used, as in the case in which the
ages of two different families, with independent calibration errors,
are to be compared.  

Among the figures, not included in this paper but available in the
Supplementary material site, there are all the V-shape plots, which
can be useful to better appreciate the robustness of our conclusions.

In this Section we also comment on ages for the same families found in
the scientific literature, with the warning that for some families
there are multiple estimates, including discordant ones, in some cases
published by the same authors at different times. Thus we think it is
important to have a source of ages computed with a uniform and well
documented procedure, such as this paper. Compilations of ages, such
as \citep{nesv05,broz13}, are useful for consultation, but have the
limitation of mixing results obtained with very different methods,
sometimes even with methods not specified. We use the terminology
\textit{consistent} when one nominal value is within the STD of the
other, \textit{compatible} when difference of nominal values is less
than the sum of the two STD, \textit{discordant} otherwise.

\subsubsection{Ages of fragmentation families}

The ages results are in Table~\ref{tab:age_frag}; comments on
specific families follow. 

\begin{table}[t!]
\footnotesize
 \centering
 \caption{Age estimation for the fragmentation families: family number
   and name, $da/dt$, age estimation, uncertainty of the age due to
   the fit, uncertainty of the age due to the calibration, and total
   uncertainty of the age estimation.}
  \label{tab:age_frag}
  \medskip
  \begin{tabular}{|l|cc|rccc|}
  \hline
number/           &side  & $da/dt$        & Age & STD(fit)& STD(cal) & STD(age) \\ 
name              &IN/OUT& $10^{-10} au/d$ & My  &   My    &   My    &   My      \\
\hline
158 Koronis       &  IN  &   -3.40        & 1792&    262   & 358 &  444     \\	
                  &  OUT &\phantom{-}3.34 & 1708&    206   & 342 &  399     \\
24  Themis        &  IN  &   -5.68        & 2447&    678   & 489 &  836     \\    
                  &  OUT &\phantom{-}5.54 & 3782&    588   & 756 &  958     \\
847 Agnia         &  IN  &   -3.46        & 1003&    207   & 201 &  288     \\
                  &  OUT &\phantom{-}3.41 &  669&    100   & 134 &  167     \\
3395 Jitka        &  IN  &   -3.44        &  136&     25   &  27 &   37     \\
                  &  OUT &\phantom{-}3.41 &  138&     24   &  28 &   37     \\
1726 Hoffmeister  &  IN  &   -5.90        &  337&     47   &  84 &   96     \\
                  &  OUT &\phantom{-}5.86 &  328&     42   &  82 &   92     \\
668 Dora          &  IN  &   -6.11        &  532&     87   & 133 &  159     \\
                  &  OUT &\phantom{-}6.08 &  471&    141   & 118 &  184     \\
434 Hungaria      &  IN  &   -3.23        &  208&     19   &  62 &   65     \\
                  &  OUT &\phantom{-}3.18 &  205&      8   &  62 &   62     \\
480 Hansa         &  IN  &   -3.53        &  763&    310   & 153 &  346     \\
                  &  OUT &\phantom{-}3.44 &  950&    117   & 190 &  223     \\
808 Merxia        &  IN  &   -3.52        &  338&     28   &  68 &   73     \\
                  &  OUT &\phantom{-}3.47 &  321&     24   &  64 &   69     \\
3330 Gantrisch    &  IN  &   -5.75        &  436&    105   & 131 &  168     \\ 
                  &  OUT &\phantom{-}5.73 &  492&    131   & 148 &  197     \\
10955 Harig \&    &  IN  &   -3.48        &  441&     78   & 132 &  154     \\
19466 Darcydiegel &  OUT &\phantom{-}3.42 &  500&    146   & 150 &  209     \\
1521 Seinajoki    &  IN  &   -3.50        &  338&     66   & 101 &  121     \\
                  &  OUT &\phantom{-}3.49 &  122&     15   &  37 &   40     \\
569 Misa          &  IN  &    -6.23       &  319&    242   &  80 &   255    \\
                  &  OUT &\phantom{-}6.15 &  249&     85   &  62 &   105    \\
15124 2000EZ$_{39}$&  IN  &   -6.22        &  111&     10   &  28 &   29     \\
                  &  OUT &\phantom{-}6.18 &  113&     11   &  28 &   30     \\
1128 Astrid       &  IN  &     -5.89      &  150&     11   &  30 &   32     \\
                  &  OUT &\phantom{-}5.85 &  150&     11   &  30 &   32     \\
845 Naema         &  IN  &   -5.73        &  149&     19   &  30 &   35     \\
                  &  OUT &\phantom{-}5.70 &  163&      8   &  33 &   34     \\
\hline
\end{tabular}
\end{table}

158 (Koronis): the present estimate increases somewhat the result we
reported in \citep{bigdata}[Table 10] of $1500$ My for the OUT side
(the result for the IN side was considered of lower quality), but
within the fit uncertainty. Now the results from the two sides are not
just consistent but very close, and the fit uncertainty has slightly
improved (in Table~10 of the previous paper the calibration
uncertainty was not included). The earliest estimates in the
literature were just upper bounds of $\leq 2$ Gy
\citep{marzari95,chapman96}, followed by
\citep{greenberg96,farinella96} who give $\sim 2$ Gy; \cite{broz13}
give $2.5\pm 1$ Gy, which is consistent with our results: our
improvement in accuracy is significant.

24 (Themis): the two sides give different values which are not
discordant, but are affected by large uncertainties. This is one of
the oldest families, for which there are few ages estimates in the
literature: \citep{marzari95} give $2$ Gy.

847 (Agnia): the new result is consistent with the one of
\citep{bigdata}[Table 10] for the IN side; the OUT side is anyway
degraded by the presence of the 3395 subfamily. 

3395 (Jitka): the results are almost the same as in
\citep{bigdata}[Table 10]. In the literature there are estimates for
the age of Agnia, in \citep{vok06c} of $100^{+30}_{-20}$ My, but from their
Figure~1 it is clear that their Agnia family is restricted to our
Jitka subfamily, apart from the addition of (847) Agnia itself. Also
in \citep{broz13} there is an estimate for 847 of $200\pm 100$ My. Thus
our results on the age are consistent with all the others, even if we
disagree on the name of the family.

1726 (Hoffmeister): our result is consistent with the one in
\citep{nesv05,broz13}, giving $300\pm 200$ My, but with significantly
lower uncertainty. The fact that the perturbations from (1) Ceres do
not appear to disturb an evolution model based on the Yarkovsky
secular drift is a confirmation of the statement by
\cite{delisle_laskar}: chaotic perturbations from other asteroids are
less effective in shifting the semimajor axis than Yarkovsky for the
objects with $D<40$ km.

668 (Dora): the OUT result is somewhat degraded by the 5/2 resonance,
thus the IN is better, but anyway they are consistent.  \citep{broz13}
give $500\pm 200$ My, in good agreement, notwithstanding the much
lower distance cutoff used to define the family ($60$ m/s vs. our $90$).

434 (Hungaria): with a similar but less rigorous method, \cite{mil10}
find $274$ My, which is higher but practically consistent with the
current result.  \cite{warner09} give $\sim 500$ My, but with a low
accuracy method.

480 (Hansa): in the literature we find only \citep{carruba10, broz13}
giving as upper bound $<1.6$ Gy.  Our result is much more informative,
especially from the OUT side.

808 (Merxia): our results are consistent with both \citep{broz13} $300\pm
200$ My, and \citep{nesv05} $500\pm 200$, but significantly more
precise.

3330 (Gantrisch): we have found nothing in the literature on the age
of this family, thus the result is useful even if the relative
accuracy is poor.

10955 (Harig), including 19466: a well determined slope, consistent
between the two sides, thus confirming the join. The absolute age is
of limited accuracy because of the lack of physical observations. No
previous estimates found in the literature.

1521 (Seinajoki) has two significantly different ages, younger for the
OUT side. This is an additional case of a dynamical family containing
two collisional families. \citep{nesv05} gives $50\pm 40$ My, which is
compatible with our estimate for the OUT side. 

1128 (Astrid) has a perfect agreement on the two sides, which appears
as a coincidence since the uncertainty is much higher. \cite{nesv05}
give $100\pm 50$ which is consistent, our estimate being more
precise.

845 (Naema) has a good agreement on the two sides. \cite{nesv05} give
$100\pm 50$ which is compatible, our estimate being more
precise.

\subsubsection{Ages of cratering families}

The ages results are in Table~\ref{tab:age_crat}; comments on
each family follow.

\begin{table}[h]
\footnotesize
 \centering
 \caption{Age estimation for the cratering families. Columns as in
   Table~\ref{tab:age_frag}.}
  \label{tab:age_crat}
  \medskip
  \begin{tabular}{|l|cc|rccc|}
  \hline
number/          &side  & $da/dt$        & Age & STD(fit) & STD(cal) & STD(age) \\ 
name             &IN/OUT& $10^{-10} au/d$ & My  &   My     &   My     &   My     \\
\hline
4 Vesta           &  IN  &    -3.60       &  930&    112   & 186 &   217    \\ 
                  &  OUT &\phantom{-}3.49 & 1906&    537   & 381 &   659    \\
15 Eunomia        &  IN  &    -3.66       & 1955&    155   & 391 &   421    \\
                  &  OUT &\phantom{-}3.55 & 1144&     57   & 229 &   236    \\
20 Massalia       &  IN  &    -3.81       &  174&      7   &  35 &    35    \\
                  &  OUT &\phantom{-}3.73 &  189&     16   &  38 &    41    \\
10 Hygiea         &  IN  &    -5.67       & 1330&    139   & 266 &   300    \\
                  &  OUT &\phantom{-}5.50 & 1368&    183   & 274 &   329    \\
31 Euphrosyne     &  IN  &    -5.71       & 1309&    169   & 262 &   312    \\
                  &  OUT &\phantom{-}5.72 & 1160&    142   & 232 &   272    \\
3 Juno            &  IN  &    -3.46       &  550&    110   & 110 &   156    \\
                  &  OUT &\phantom{-}3.41 &  370&    143   &  74 &   161    \\
163 Erigone \&    &  IN  &    -6.68       &  212&     53   &  42 &    68    \\
5026 Martes       &  OUT &\phantom{-}6.64 &  230&     46   &  19 &    50    \\
\hline
\end{tabular}
\end{table}

4 (Vesta): the idea that Vesta might have suffered two large impacts
generating two families \citep{bigdata}[Sec. 7.3] is quite natural
given that cratering does not decrease the collisional cross section,
and has been proposed long ago \citep{farinella96}. The new error
model and outlier rejection procedure have reduced the fit
uncertainty, especially for the OUT side, thus the ratio of values on
the two sides has increased its level of significance (see
Table~\ref{tab:slope_crater}). The good agreement of the age from the
IN side with the cratering age of the Rheasilvia basin, $1$ Gy
according to \cite{marchi12} is very interesting. Only a rough lower
bound age of $\sim 2$ Gy is available for the Veneneia basin because
of the disruption due to the impact forming Rheasilvia
\citep{obrien14}. Thus our age estimate from the OUT side is an
independent constraint to the age of Veneneia.

15 (Eunomia): in \citep{bigdata}[Table 10] the difference in the slopes
for the two sides was much smaller and the fit uncertainty for the OUT
side much larger, thus the existence of two separate ages was
proposed as possible. The improved results provide a ratio very
significantly different from 1, thus the existence of two collisional
families inside the single dynamical family 15 is now supported by
high S/N evidence.  \cite{nesv05} give $2.5\pm 0.5$ Gy as age for the
entire family, which is compatible with our IN side age.

20 (Massalia): our new results are very similar to the ones of our
previous paper as well as consistent with \citep{vok06b}, giving as
most likely an age between $150$ and $200$ My. On the contrary
\citep{nesv03} give $300\pm 100$ which is marginally compatible.

10 (Hygiea): the interesting point is that this dynamical family
appears to have a single age, a non-trivial result since the family
has a bimodal shape in the proper $(a,e)$ projection, and (10) has
almost the same impact cross section as (4) Vesta. In the literature
we found only \citep{nesv05} giving a consistent, but low accuracy,
$2\pm 1$ Gy.

31 (Euphrosine): This high proper $\sin{I}$ family is crossed by many
resonances, nevertheless the age can be estimated. In the
literature, we found only the upper bound $<1.5$ Gy in \citep{broz13}.

3 (Juno): the two ages IN and OUT are not consistent but only
compatible; more data are needed to assess the possibility of multiple
collisions. In the literature we found only an upper bound $<700$ My
in \citep{broz13}.

163 (Erigone): another very good example of join of two dynamical
families, 163 and 5026, into a collisional family with all the
properties expected, including age estimates consistent (within half
of STD) and a lower number density in a central strip. \citep{vok06b}
give an age of $280\pm 112$ My, which is higher but consistent;
\citep{bot15} by a different method give an age $170^{+25}_{-30}$, which is
lower but consistent with the IN side. From the figures we can deduce
that in both papers their family 163 also includes our 5026.

\subsubsection{Ages of young families}
\label{s:ages_young}

The ages results are in Table~\ref{tab:age_young}. We are interested
in finding a lower limit for the ages we can compute with the V-shape
method. For most of these asteroids there are in the literature only
either upper bounds or low relative accuracy estimates of the ages
\citep{broz13}. In order of estimated age:

\begin{table}[t!]
\footnotesize
 \centering
 \caption{Age estimation for the young families. Columns as in
   Table~\ref{tab:age_frag}.}
  \label{tab:age_young}
  \medskip
  \begin{tabular}{|l|cc|rccc|}
  \hline
number/          &side  & $da/dt$        &  Age & STD(fit)& STD(cal) &STD(age) \\ 
name             &IN/OUT& $10^{-10} au/d$ & My  &   My     &   My     & My     \\
\hline
3815 K\"onig     &  IN  &     -6.21      &    51&       6  &  13 &    14    \\
                 &  OUT &\phantom{-}6.21 &    51&       6  &  13 &    14    \\
396 Aeolia       &  IN  &     -3.09      &   100&      18  &  25 &    31    \\
                 &  OUT &\phantom{-}3.08 &    91&      15  &  23 &    27    \\
606 Brangane     &  IN  &     -3.82      &    48&       4  &  10 &    10    \\
                 &  OUT &\phantom{-}3.81 &    44&       7  &   9 &    11    \\
1547 Nele        &  IN  &      -3.61     &    14&       2  &   4 &     5    \\
     	         &  OUT &\phantom{-}3.61 &    15&       5  &   5 &     7    \\
18405 1993FY$_{12}$& IN &       -3.50     &    83&      28  &  25 &    37   \\
     	         &  OUT &\phantom{-}3.48 &    83&       13 &  25 &    28    \\
\hline
\end{tabular}
\end{table}

1547 (Nele): for this family \cite{broz13} give an age $<40$ My;
\cite{nesv03} give a constraint $\leq 5$ My on the age of the
Iannini cluster, which he identified as composed of $18$ members not
including (1547). Our estimate (for a family with $152-3=149$ members,
including (4652) Iannini) is higher, but such a young age could be too
much affected by the effect of the initial velocity field, which is
apparent in the anti-correlation between proper $a,e$. From this
example we conclude that probably $15$ My is too young to be an
accurate estimate by the V-shape method; this family should be
dated by a method using also the evolution of the angles $\varpi,
\Omega$.

3815 (K\"onig): we have a precise estimate, in the literature we have
found only an upper bound $<100$ My \citep{broz13}. 

606 (Brangane): also a precise estimate, in good agreement with $50\pm
40$ in \citep{broz13}.  We do not have a ground truth to assess the
systematic error due to contamination from the initial velocity
spread, which for these ages may not be negligible\footnote{A size
  independent velocity spread is removed by our fit method, but there
  may well be a $1/D$ dependency in this spread.}.

396 (Aeolia): also a precise estimate, consistent with the upper bound
$<100$ My in \citep{broz13}. 
18405 (1993FY$_{12}$): \cite{broz13} give an age $< 200$ My. Our
estimate is precise and not just consistent, but the same on the two
sides.  For this range of ages around $100$ My the initial velocity
field should not matter.

From these examples we can conclude that the V-shape method is
applicable to \textit{young} families with ages below $100$ My, but
there is some lower age limit $t_{min}$ such that younger ages are
inaccurately estimated from the V-shape.  The cases we have analyzed
suggest that $t_{min}> 15$ My, but we do not have enough information
to set an upper bound for $t_{min}$.

\subsubsection{Ages of one-sided families}

The ages results are in Table~\ref{tab:age_one}; these ages are based
upon the assumption that only one side of the family V-shape is
preserved.  Of course if this was not the case, ages younger by factor
roughly 2 would be obtained. For each case, comments on the
justification of the one-side assumption are given below.

\begin{table}[h!]
\footnotesize
 \centering
 \caption{Age estimation for the one-sided families. Columns as in
   Table~\ref{tab:age_frag}.}
  \label{tab:age_one}
  \medskip
  \begin{tabular}{|l|cc|rccc|}
  \hline
number/          &side  & $da/dt$        & Age & STD(fit)  & STD(cal) & STD(age) \\ 
name             &IN/OUT& $10^{-10} au/d$ & My  &   My      &   My    & My      \\
\hline
170 Maria        &  OUT &\phantom{-}3.48 &  1932&   169     & 386 &   422     \\
1272 Gefion      &  IN  &     -3.50      &  1103&   270     & 276 &   386     \\
2076 Levin       &  OUT &\phantom{-}3.86 &   366&    59     & 110 &   125     \\
3827 Zdenekhorsky&  IN  &     -5.99      &   154&    14     &  31 &    34     \\
1658 Innes       &  OUT &\phantom{-}3.59 &   464&    31     & 139 &   143     \\
375 Ursula       &  IN  &     -5.56      &  3483&   765     & 697 &  1035     \\
\hline
\end{tabular}
\end{table}

170 (Maria): the very strong 3/1 resonance with Jupiter makes it
impossible for asteroids of the IN side of the family to have survived
in the main belt, moreover the shape of the family in the $(a, 1/D)$
plane is unequivocally one sided. This is an ancient family, and our
age estimate is compatible with $3\pm 1$ Gy given in \citep{nesv05},
but we have significantly decreased the estimate, to the point that
this cannot be a ``LHB'' family, as suggested by \citep{broz13}.

1272 (Gefion): the very strong 5/2 resonance with Jupiter makes it
impossible for most asteroids of the OUT side of the family to have
survived in the main belt. Thus there is no OUT side in the
V-shape\footnote{See the figure 1272\_vshapea.eps in the Supplementary
  material.}.  \cite{nesv05} give an age $1.2\pm 0.4$ Gy, in good
agreement with ours, while \citep{nesv09}[Figure 1] show a one-sided
model, giving a discordant age of $480\pm50$ My.

2076 (Levin): as discussed in Section~\ref{s:results}, this could be
just a component of a complex family, possibly including 298 and 883.
The OUT slope, thus the age we have estimated, refers to the event
generating 2076, while 298 and 883 have too few members for a reliable
age. In the literature there are ages for the family of (298)
Baptistina: e.g., \cite{botk07nat} give a discordant age of
$160^{+30}_{-20}$ My, but they refer to a two-sided V-shape including
our 883, with an enormous number of outliers.

3827 (Zdenekhorsky): the family shape is obviously asymmetric, with
much fewer members on the OUT side\footnote{See the figure
  3827\_vshapea.eps in the supplementary material.}. This prevents a
statistically significant determination of the OUT slope. The family
is not abruptly truncated, possibly because the effect of (1) Ceres is
weaker than the one of the main resonances with Jupiter.

1658 (Innes): the shape of the family in the $(a, 1/D)$ plane is
clearly one-sided. The family ends on the IN side a bit too
far from the 3/1 resonance, thus the dynamics of the depletion on the
IN side remains to be investigated.

375 (Ursula): the strongest 1/2 resonance with Jupiter makes it
impossible for most asteroids of the OUT side of the family to have
survived in the main belt. This prevents a statistically significant
determination of the OUT slope. With an age estimated at $\sim 3.5\pm
1$ Gy, this family could be the oldest for which we have an
age. \cite{broz13} give the upper bound $<3.5$ Gy. 

\section{Conclusions and future work}

In this paper we have computed the ages of $37$ collisional
families\footnote{Plus one possible, a second age for the family 3
  (Juno) with a moderate significance in the slope ratio, see
  Table~\ref{tab:slope_crater}.}. The members of these collisional
families belong to $34$ dynamical families, including $30$ of those
with $>250$ members. Moreover, we have computed uncertainties based on
a well defined error model: the standard deviations for the ages are
quite large in many cases, but still the signal to noise ratio is
significantly $>1$.

\subsection{Main results}

In Figure~\ref{fig:famage_plotnf} we have placed the families on the
horizontal axis with the same order used in the Tables, separated in
four categories\footnote{To locate these families in the asteroid
  belt, the best way is to use the graphic visualizer of asteroid
  families provided by the AstDyS site at
  http://hamilton.dm.unipi.it/astdys2/Plot/}.

On the vertical axis (in a logarithmic scale) we have
marked the estimated age with a 1 STD error bar. To avoid overcrowding
of the Figure, for the families with compatible ages from the IN and
OUT side we have used the average (weighted with the inverse square of
the STD) as the nominal with an error bar $\sigma=
\sqrt{\sigma_{IN}^2+\sigma_{OUT}^2}/2$. If the two ages are
incompatible we have plotted the two estimates with the corresponding
bars\footnote{For 847 we have used the IN age and STD, as discussed in
  Section~\ref{s:results}.}. We have also used an informal terminology
by which families are rated by their age: \textit{primordial} with age
$>3.7$ Gy, \textit{ancient} with age between $1$ and $3.7$ Gy,
\textit{old} with age between $0.1$ and $1$ Gy, and finally the
adjective \textit{young}, as used previously, is for ages $<0.1$ Gy.

\begin{figure}[h!]
\figfigincl{14 cm}{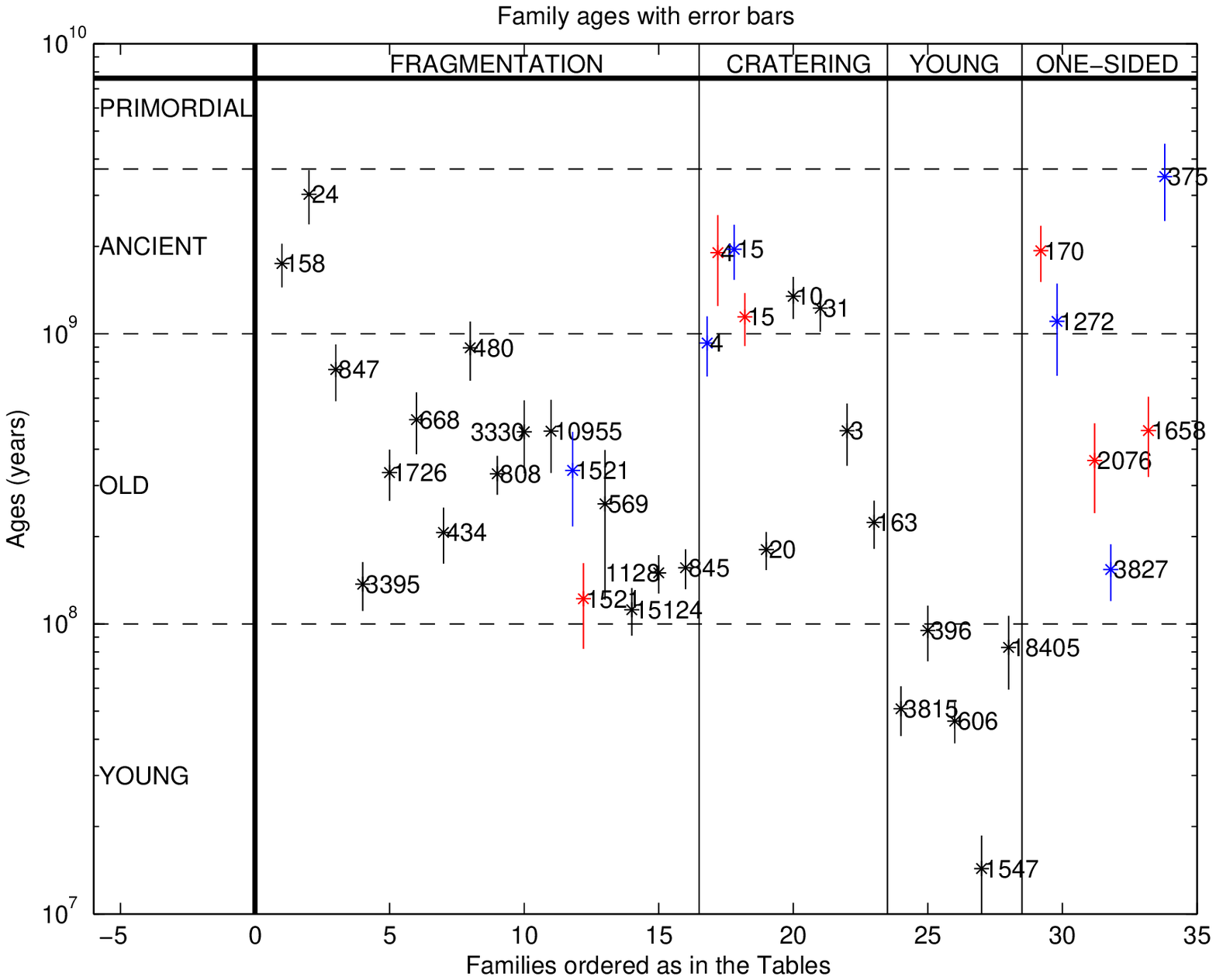}{Family ages and their uncertainties
  computed in this paper. If two ages of the same family are
  incompatible, the figure shows both ages; this applies to
    families 4, 15, 1521. The horizontal dashed lines separate the
  conventional age groupings, the vertical solid lines separate the
  family types, for the definitions see in the text.}
\end{figure}

By looking at Figure~\ref{fig:famage_plotnf} it is apparent that we
have been quite successful in computing ages for old families, we have
significant results for both young and ancient, while we have little,
if any, evidence for primordial families. This should not be rated as
a surprise: already \cite{broz13}, while specifically searching for
primordial families, found a very short list of candidates, out of
which 4, 10, 15, 158 and 170 we are showing to be ancient, but not
primordial. From our results, only two families could be primordial,
24 and 375, although they are more likely to be just ancient. Thus we
agree with the conclusion by \cite{vok10} that most of the primordial
families, which undoubtedly have existed, have been depleted of
members to the point of not being recognized by statistically
significant number density contrast: our results indicate that this
conclusion applies not only to the Cybele region (beyond the 2/1
resonance) but to the entire main belt.

Figure~\ref{fig:famage_plotnf} also shows that our results allow many
statistically significant absolute age comparisons between different
families. Although the results should be improved, especially by
obtaining more accurate Yarkovsky calibrations, this can be the
beginning of a real asteroid belt chronology. The large compilations
of family ages, such as \cite{nesv05, broz13} are very useful to
confirm that our results are reasonable. When available, the
uncertainties reported in these compilations are generally larger;
often only upper/lower bounds are given.  However, the literature as
analyzed in Section~\ref{s:ages} shows that often results obtained
with different methods, even by the same authors, can be discordant.
Thus the comparison of ages for different families should not be done
with the ages listed in a compilation, but only from a list of ages
computed with a single consistent method, including a single
consistent calibration scheme, as in this paper.

In the previous paper \cite{bigdata} we had introduced the distinction
between dynamical and collisional families; out of the 5 dynamical
families we analyzed as examples, we found 3 cases in which a
dynamical family corresponds to at least 2 collisional ones.  In this
paper we report on the results of a systematic survey of the largest
(by number of members) dynamical families, monitoring whether the 1 to
1 correspondence with collisional families does or does not apply.

We have found two examples, for which we use the definition of
\textit{family join}, in which two separate dynamical families
together form a single V-shape, with consistent slopes, thus
indicating a single collisional event: this applies to families 10955
and 19466, 163 and 5026. Note that this is distinct from a
\textit{family merge} which can arise when two families, as a result
of adding new members with recently computed proper elements, acquire
some members in common \citep{namur_update}.

We have also found at least three examples of dynamical families
containing multiple collisional families: 4, 15 and 1521. For these we
have obtained discordant slopes from the IN and the OUT side of the
V-shape, resulting in distinct ages, see Figure~\ref{fig:famage_plotnf}.
We have found a dubious case, family 3, and there are several other
cases already either known or suspected.

Finally, we have found two cases of families containing a conspicuous
subfamily, with a sharp number density contrast, such that it is
possible to measure the slope of a distinct V-shape for the subfamily,
thus the age of the secondary collision: the subfamily 3395 of 847,
and 15124 of 569. There are several cases of subfamilies, with a
separate collisional age, already reported in the literature, but they
are mostly from recent ($<10$ My of age) collisions: we have
identified subfamilies with ages of $\sim 100$ My.

From the above discussion, we think a new paradigm emerges: whenever a
family age computation is performed, the question on the minimum
number of collisional events capable of generating the observed
distribution of members of the family in the classification space has
to be analyzed.  This needs to take also into account other families
in the neighborhood (in the classification space). In our case, the
classification space is the 3-dimensional proper elements space
because we use dynamical families, but note that the same argument
applies also to other classifications made in different spaces, such
as the ones containing also physical observations data: separate
collisional families may well have the same composition.

\subsection{Open problems}
\label{s:open}

On other issues we have accumulated data, useful to constrain the
asteroid families evolution, but we do not have a full model.

An example is the fact already known that many families have a central
gap, in the sense of a bimodal number frequency distribution of
members as a function of proper $a$. The interpretation of this gap as
a consequence of the interaction between the YORP and the Yarkovsky
effect, as proposed in \citep{vok06b}, is plausible and widely
accepted, but a model capable of predicting the timescales of this
evolution is not available.

We have observed the presence and depth of the gap for all the
families having, in our best estimate, $<600$ My. 

\begin{itemize}

\item Ages between $10$ and $100$ My: the gap does not occur
  in the youngest 1547 and the one near the upper limit of $ 100$ My,
  that is 396, but occurs in 18405 which has an age similar to 396,
  and in the two with ages $\sim 50$ My, 3815 and 606.

\item Ages between $100$ and $200$ My: the gap occurs
  consistently in families such as 3395, 15124, 1128, 845, and less
  deep in 20.

\item Ages between $200$ and $400$ My: there are three
  families with gap (434, 808, 163) and two without (1726, 3).

\item Ages between $400$ and $600$ My: 10955 has a gap and 668 does
  not.

\item Ages $>600$ My: among the ancient families only 158 and
  maybe 31 show some small dip in density at the center.

\end{itemize}

These results do not contradict the interpretation that YORP moves the
rotation axes towards the spin up/spin down position, but takes quite
some time to achieve a strong bimodality which gradually empties the
gap.  Over longer time scales, spin axis randomization can reverse the
process.  However, our set of examples above shows that the time
scales for such processes are not uniform, but may substantially
change from family to family.

Another open problem results from the fact that several families on
the outer edge of the 3/1 resonance gap appear to have a boundary
close to, but not at the Kirkwood gap. This happens to the IN side of
families 480 and 15; there are also families 170 and 1658 which are
one-sided because of the missing IN side, with the family not touching
the gap.  This might require a dedicated study to find a plausible
explanation.

\subsection{Family ages left to be computed}

Of the dynamical families in the current classification, there are
$11$ with $>300$ members for which we have not yet computed a
satisfactory age. The motivations are as follows.

\begin{itemize}

\item There are five complex families: 135, known to have at least two
  collisional families, with incompatible physical properties,
  difficult to disentangle; see e.g., \citep{bigdata}[Figure 10]; 221,
  complex both for dynamical evolution \citep{vok06a} and suspect of
  multiple collisions; 145, which appears to have at least 2 ages;
  25, corresponding to a stable region surrounded by secular
  resonances, could have many collisional families; 179, a cratering
  family which is difficult to be interpreted.

\item There are another four families strongly affected in their shape in
  proper element space by resonances: 5, 110, 283 with
  secular resonances, and 1911 inside the 3/2 resonance. 

 \item Two others: 490, well known to be of recent age
   \citep{nesv03,tsiganis}; 1040, at large proper $\sin{I}$ and also
   quite large $e$; both are strongly affected by 3-body resonances.

\end{itemize}

We are convinced that for many of these it will be possible to
estimate the age, but this might require ad hoc methods, different
from case to case. In this paper we have included all the ages which
we have up to now been able to estimate by a uniform method.

Other families with marginal number of members for the V-shape fit
(between 100 and 300 in the current classification) could become
suitable as new proper elements are computed and the classification is
automatically updated, especially in the zones where the number
density is low, such as the high $I$ region, and the Cybele region,
beyond the 2/1 resonance.

\section*{Acknowledgments}

The authors have been supported for this research by: ITN Marie Curie
``STARDUST -- the Asteroid and Space Debris Network''
(FP7-PEOPLE-2012-ITN, Project number 317185) (A.M. and Z.K), the
Department of Mathematics of the University of Pisa (A.M. and F.S.),
SpaceDyS srl, Cascina, Italy (F.S.), and the Ministry of Education,
Science and Technological Development of Serbia, under the project
176011 (Z.K.).

The authors would also like to thank Melissa Dykhuis and Marco Delb\`o
for their helpful comments.

\section*{References}

\bibliographystyle{elsarticle-harv}
\bibliography{family_ages_biblio}

\begin{thebibliography}{33}
\expandafter\ifx\csname natexlab\endcsname\relax\def\natexlab#1{#1}\fi
\expandafter\ifx\csname url\endcsname\relax
  \def\url#1{\texttt{#1}}\fi
\expandafter\ifx\csname urlprefix\endcsname\relax\def\urlprefix{URL }\fi

\bibitem[{{Bottke} et~al.(2007){Bottke}, {Vokrouhlick{\'y}}, and
  {Nesvorn{\'y}}}]{botk07nat}
{Bottke}, W.~F., {Vokrouhlick{\'y}}, D., {Nesvorn{\'y}}, D., Sep. 2007. {An
  asteroid breakup 160Myr ago as the probable source of the K/T impactor}.
  Nature 449, 48--53.

\bibitem[{{Bottke} et~al.(2015){Bottke}, {Vokrouhlick{\'y}}, {Walsh},
  {Delb\`o}, {Michel}, {Lauretta}, {Campins}, {Connolly}, {Scheeres}, and
  {Chesley}}]{bot15}
{Bottke}, W.~F., {Vokrouhlick{\'y}}, D., {Walsh}, K., {Delb\`o}, M., {Michel},
  P., {Lauretta}, D., {Campins}, H., {Connolly}, H.~J., {Scheeres}, D.,
  {Chesley}, S., Jan. 2015. {In search of the source of asteroid (101955)
  Bennu: Applications of the stochastic YORP model}. Icarus 245, 191--217.

\bibitem[{{Bro{\v z}} et~al.(2013){Bro{\v z}}, {Morbidelli}, {Bottke},
  {Rozehnal}, {Vokrouhlick{\'y}}, and {Nesvorn{\'y}}}]{broz13}
{Bro{\v z}}, M., {Morbidelli}, A., {Bottke}, W.~F., {Rozehnal}, J.,
  {Vokrouhlick{\'y}}, D., {Nesvorn{\'y}}, D., Mar. 2013. {Constraining the
  cometary flux through the asteroid belt during the late heavy bombardment}.
  A\&A 551, A117.

\bibitem[{{Carpino} et~al.(2003){Carpino}, {Milani}, and {Chesley}}]{carpino03}
{Carpino}, M., {Milani}, A., {Chesley}, S.~R., Dec. 2003. {Error statistics of
  asteroid optical astrometric observations}. Icarus 166, 248--270.

\bibitem[{{Carruba}(2010)}]{carruba10}
{Carruba}, V., Oct. 2010. {The stable archipelago in the region of the Pallas
  and Hansa dynamical families}. MNRAS 408, 580--600.

\bibitem[{{Carry}(2012)}]{carry12}
{Carry}, B., 2012. {Density of asteroids}. Planetary and Space Science 73,
  98--118.

\bibitem[{{Chapman} et~al.(1996){Chapman}, {Ryan}, {Merline}, {Neukum},
  {Wagner}, {Thomas}, {Veverka}, and {Sullivan}}]{chapman96}
{Chapman}, C.~R., {Ryan}, E.~V., {Merline}, W.~J., {Neukum}, G., {Wagner}, R.,
  {Thomas}, P.~C., {Veverka}, J., {Sullivan}, R.~J., Mar. 1996. {Cratering on
  Ida}. Icarus 120, 77--86.

\bibitem[{{Chesley} et~al.(2014){Chesley}, {Farnocchia}, {Nolan},
  {Vokrouhlick{\'y}}, {Chodas}, {Milani}, {Spoto}, {Rozitis}, {Benner},
  {Bottke}, {Busch}, {Emery}, {Howell}, {Lauretta}, {Margot}, and
  {Taylor}}]{bennu}
{Chesley}, S.~R., {Farnocchia}, D., {Nolan}, M.~C., {Vokrouhlick{\'y}}, D.,
  {Chodas}, P.~W., {Milani}, A., {Spoto}, F., {Rozitis}, B., {Benner},
  L.~A.~M., {Bottke}, W.~F., {Busch}, M.~W., {Emery}, J.~P., {Howell}, E.~S.,
  {Lauretta}, D.~S., {Margot}, J.-L., {Taylor}, P.~A., Jun. 2014. {Orbit and
  bulk density of the OSIRIS-REx target Asteroid (101955) Bennu}. Icarus 235,
  5--22.

\bibitem[{{Delisle} and {Laskar}(2012)}]{delisle_laskar}
{Delisle}, J.-B., {Laskar}, J., Apr. 2012. {Chaotic diffusion of the Vesta
  family induced by close encounters with massive asteroids}. A\&A 540, A118.

\bibitem[{{Farinella} et~al.(1996){Farinella}, {Davis}, and
  {Marzari}}]{farinella96}
{Farinella}, P., {Davis}, D.~R., {Marzari}, F., 1996. {Asteroid Families, Old
  and Young}. In: {Rettig}, T., {Hahn}, J.~M. (Eds.), Completing the Inventory
  of the Solar System. Vol. 107 of Astronomical Society of the Pacific
  Conference Series. pp. 45--55.

\bibitem[{{Greenberg} et~al.(1996){Greenberg}, {Bottke}, {Nolan}, {Geissler},
  {Petit}, {Durda}, {Asphaug}, and {Head}}]{greenberg96}
{Greenberg}, R., {Bottke}, W.~F., {Nolan}, M., {Geissler}, P., {Petit}, J.-M.,
  {Durda}, D.~D., {Asphaug}, E., {Head}, J., Mar. 1996. {Collisional and
  Dynamical History of Ida}. Icarus 120, 106--118.

\bibitem[{{Kne\v zevi\'c} et~al.(2014){Kne\v zevi\'c}, {Milani}, {Cellino},
  {Novakovi{\'c}}, {Spoto}, and {Paolicchi}}]{namur_update}
{Kne\v zevi\'c}, Z., {Milani}, A., {Cellino}, A., {Novakovi{\'c}}, B., {Spoto},
  F., {Paolicchi}, P., 2014. {Automated classification of asteroids into
  families at work}. In: {Kne\v zevi\'c}, Z., {Lemaitre}, A. (Eds.), Complex
  Planetary Systems. Proceedings of the IAU Symposia, Cambridge Univ. Press.
  pp. 130--133.

\bibitem[{{Mainzer} et~al.(2011){Mainzer}, {Grav}, {Bauer}, {Masiero},
  {McMillan}, {Cutri}, {Walker}, {Wright}, {Eisenhardt}, {Tholen}, {Spahr},
  {Jedicke}, {Denneau}, {DeBaun}, {Elsbury}, {Gautier}, {Gomillion}, {Hand},
  {Mo}, {Watkins}, {Wilkins}, {Bryngelson}, {Del Pino Molina}, {Desai},
  {G{\'o}mez Camus}, {Hidalgo}, {Konstantopoulos}, {Larsen}, {Maleszewski},
  {Malkan}, {Mauduit}, {Mullan}, {Olszewski}, {Pforr}, {Saro}, {Scotti}, and
  {Wasserman}}]{Mainzer_WISE}
{Mainzer}, A., {Grav}, T., {Bauer}, J., {Masiero}, J., {McMillan}, R.~S.,
  {Cutri}, R.~M., {Walker}, R., {Wright}, E., {Eisenhardt}, P., {Tholen},
  D.~J., {Spahr}, T., {Jedicke}, R., {Denneau}, L., {DeBaun}, E., {Elsbury},
  D., {Gautier}, T., {Gomillion}, S., {Hand}, E., {Mo}, W., {Watkins}, J.,
  {Wilkins}, A., {Bryngelson}, G.~L., {Del Pino Molina}, A., {Desai}, S.,
  {G{\'o}mez Camus}, M., {Hidalgo}, S.~L., {Konstantopoulos}, I., {Larsen},
  J.~A., {Maleszewski}, C., {Malkan}, M.~A., {Mauduit}, J.-C., {Mullan}, B.~L.,
  {Olszewski}, E.~W., {Pforr}, J., {Saro}, A., {Scotti}, J.~V., {Wasserman},
  L.~H., Dec. 2011. {NEOWISE Observations of Near-Earth Objects: Preliminary
  Results}. The Astrophysical Journal 743, 156.

\bibitem[{{Marchi} et~al.(2012){Marchi}, {McSween}, {O'Brien}, {Schenk}, {De
  Sanctis}, {Gaskell}, {Jaumann}, {Mottola}, {Preusker}, {Raymond}, {Roatsch},
  and {Russell}}]{marchi12}
{Marchi}, S., {McSween}, H.~Y., {O'Brien}, D.~P., {Schenk}, P., {De Sanctis},
  M.~C., {Gaskell}, R., {Jaumann}, R., {Mottola}, S., {Preusker}, F.,
  {Raymond}, C.~A., {Roatsch}, T., {Russell}, C.~T., May 2012. {The Violent
  Collisional History of Asteroid 4 Vesta}. Science 336, 690--694.

\bibitem[{{Marzari} et~al.(1995){Marzari}, {Davis}, and {Vanzani}}]{marzari95}
{Marzari}, F., {Davis}, D., {Vanzani}, V., Jan. 1995. {Collisional evolution of
  asteroid families}. Icarus 113, 168--187.

\bibitem[{{Masiero} et~al.(2011){Masiero}, {Mainzer}, {Grav}, {Bauer}, {Cutri},
  {Dailey}, {Eisenhardt}, {McMillan}, {Spahr}, {Skrutskie}, {Tholen}, {Walker},
  {Wright}, {DeBaun}, {Elsbury}, {Gautier}, {Gomillion}, and
  {Wilkins}}]{Masiero_WISE}
{Masiero}, J.~R., {Mainzer}, A.~K., {Grav}, T., {Bauer}, J.~M., {Cutri}, R.~M.,
  {Dailey}, J., {Eisenhardt}, P.~R.~M., {McMillan}, R.~S., {Spahr}, T.~B.,
  {Skrutskie}, M.~F., {Tholen}, D., {Walker}, R.~G., {Wright}, E.~L., {DeBaun},
  E., {Elsbury}, D., {Gautier}, IV, T., {Gomillion}, S., {Wilkins}, A., Nov.
  2011. {Main Belt Asteroids with WISE/NEOWISE. I. Preliminary Albedos and
  Diameters}. The Astrophysical Journal 741, 68.

\bibitem[{{Milani} et~al.(2014){Milani}, {Cellino}, {Kne{\v z}evi{\'c}},
  {Novakovi{\'c}}, {Spoto}, and {Paolicchi}}]{bigdata}
{Milani}, A., {Cellino}, A., {Kne{\v z}evi{\'c}}, Z., {Novakovi{\'c}}, B.,
  {Spoto}, F., {Paolicchi}, P., 2014. {Asteroid families classification:
  Exploiting very large datasets}. Icarus 239, 46--73.

\bibitem[{{Milani} et~al.(2010){Milani}, {Kne{\v z}evi{\'c}}, {Novakovi{\'c}},
  and {Cellino}}]{mil10}
{Milani}, A., {Kne{\v z}evi{\'c}}, Z., {Novakovi{\'c}}, B., {Cellino}, A., Jun.
  2010. {Dynamics of the Hungaria asteroids}. Icarus 207, 769--794.

\bibitem[{{Nesvorn{\'y}} et~al.(2003){Nesvorn{\'y}}, {Bottke}, {Levison}, and
  {Dones}}]{nesv03}
{Nesvorn{\'y}}, D., {Bottke}, W.~F., {Levison}, H.~F., {Dones}, L., Jul. 2003.
  {Recent Origin of the Solar System Dust Bands}. Astrophys.J 591, 486--497.

\bibitem[{{Nesvorn{\'y}} et~al.(2005){Nesvorn{\'y}}, {Jedicke}, {Whiteley}, and
  {Ivezi{\'c}}}]{nesv05}
{Nesvorn{\'y}}, D., {Jedicke}, R., {Whiteley}, R.~J., {Ivezi{\'c}}, {\v Z}.,
  Jan. 2005. {Evidence for asteroid space weathering from the Sloan Digital Sky
  Survey}. Icarus 173, 132--152.

\bibitem[{{Nesvorn{\'y}} et~al.(2009){Nesvorn{\'y}}, {Vokrouhlick{\'y}},
  {Morbidelli}, and {Bottke}}]{nesv09}
{Nesvorn{\'y}}, D., {Vokrouhlick{\'y}}, D., {Morbidelli}, A., {Bottke}, W.~F.,
  Apr. 2009. {Asteroidal source of L chondrite meteorites}. Icarus 200,
  698--701.

\bibitem[{{O'Brien} et~al.(2014){O'Brien}, {Marchi}, {Morbidelli}, {Bottke},
  {Schenk}, {Russell}, and {Raymond}}]{obrien14}
{O'Brien}, D.~P., {Marchi}, S., {Morbidelli}, A., {Bottke}, W.~F., {Schenk},
  P.~M., {Russell}, C.~T., {Raymond}, C.~A., Nov. 2014. {Constraining the
  cratering chronology of Vesta}. Planetary Space Sci. 103, 131--142.

\bibitem[{{Pravec} et~al.(2012){Pravec}, {Harris}, {Ku{\v s}nir{\'a}k},
  {Gal{\'a}d}, and {Hornoch}}]{pravecharris}
{Pravec}, P., {Harris}, A.~W., {Ku{\v s}nir{\'a}k}, P., {Gal{\'a}d}, A.,
  {Hornoch}, K., Sep. 2012. {Absolute magnitudes of asteroids and a revision of
  asteroid albedo estimates from WISE thermal observations}. Icarus 221,
  365--387.

\bibitem[{{Shepard} et~al.(2008){Shepard}, {Kressler}, {Clark}, {Ockert-Bell},
  {Nolan}, {Howell}, {Magri}, {Giorgini}, {Benner}, and {Ostro}}]{shepard08}
{Shepard}, M.~K., {Kressler}, K.~M., {Clark}, B.~E., {Ockert-Bell}, M.~E.,
  {Nolan}, M.~C., {Howell}, E.~S., {Magri}, C., {Giorgini}, J.~D., {Benner},
  L.~A.~M., {Ostro}, S.~J., May 2008. {Radar observations of E-class Asteroids
  44 Nysa and 434 Hungaria}. Icarus 195, 220--225.

\bibitem[{{Tsiganis} et~al.(2007){Tsiganis}, {Kne{\v z}evi{\'c}}, and
  {Varvoglis}}]{tsiganis}
{Tsiganis}, K., {Kne{\v z}evi{\'c}}, Z., {Varvoglis}, H., Feb. 2007.
  {Reconstructing the orbital history of the Veritas family}. Icarus 186,
  484--497.

\bibitem[{{Vokrouhlick{\'y}} et~al.(2006{\natexlab{a}}){Vokrouhlick{\'y}},
  {Bro{\v z}}, {Bottke}, {Nesvorn{\'y}}, and {Morbidelli}}]{vok06c}
{Vokrouhlick{\'y}}, D., {Bro{\v z}}, M., {Bottke}, W.~F., {Nesvorn{\'y}}, D.,
  {Morbidelli}, A., Aug. 2006{\natexlab{a}}. {The peculiar case of the Agnia
  asteroid family}. Icarus 183, 349--361.

\bibitem[{{Vokrouhlick{\'y}} et~al.(2006{\natexlab{b}}){Vokrouhlick{\'y}},
  {Bro{\v z}}, {Bottke}, {Nesvorn{\'y}}, and {Morbidelli}}]{vok06b}
{Vokrouhlick{\'y}}, D., {Bro{\v z}}, M., {Bottke}, W.~F., {Nesvorn{\'y}}, D.,
  {Morbidelli}, A., May 2006{\natexlab{b}}. {Yarkovsky/YORP chronology of
  asteroid families}. Icarus 182, 118--142.

\bibitem[{{Vokrouhlick{\'y}} et~al.(2006{\natexlab{c}}){Vokrouhlick{\'y}},
  {Bro{\v z}}, {Morbidelli}, {Bottke}, {Nesvorn{\'y}}, {Lazzaro}, and
  {Rivkin}}]{vok06a}
{Vokrouhlick{\'y}}, D., {Bro{\v z}}, M., {Morbidelli}, A., {Bottke}, W.~F.,
  {Nesvorn{\'y}}, D., {Lazzaro}, D., {Rivkin}, A.~S., May 2006{\natexlab{c}}.
  {Yarkovsky footprints in the Eos family}. Icarus 182, 92--117.

\bibitem[{{Vokrouhlick{\'y}} et~al.(2000){Vokrouhlick{\'y}}, {Milani}, and
  {Chesley}}]{vok_2000}
{Vokrouhlick{\'y}}, D., {Milani}, A., {Chesley}, S.~R., 2000. {Yarkovsky Effect
  on Small Near-Earth Asteroids: Mathematical Formulation and Examples}. Icarus
  148, 118--138.

\bibitem[{{Vokrouhlick{\'y}} et~al.(2010){Vokrouhlick{\'y}}, {Nesvorn{\'y}},
  {Bottke}, and {Morbidelli}}]{vok10}
{Vokrouhlick{\'y}}, D., {Nesvorn{\'y}}, D., {Bottke}, W.~F., {Morbidelli}, A.,
  Jun. 2010. {Collisionally Born Family About 87 Sylvia}. Astron.J. 139,
  2148--2158.

\bibitem[{{Warner} et~al.(2009){Warner}, {Harris}, {Vokrouhlick{\'y}},
  {Nesvorn{\'y}}, and {Bottke}}]{warner09}
{Warner}, B.~D., {Harris}, A.~W., {Vokrouhlick{\'y}}, D., {Nesvorn{\'y}}, D.,
  {Bottke}, W.~F., Nov. 2009. {Analysis of the Hungaria asteroid population}.
  Icarus 204, 172--182.

\bibitem[{{Wright} et~al.(2010){Wright}, {Eisenhardt}, {Mainzer}, {Ressler},
  {Cutri}, {Jarrett}, {Kirkpatrick}, {Padgett}, {McMillan}, {Skrutskie},
  {Stanford}, {Cohen}, {Walker}, {Mather}, {Leisawitz}, {Gautier}, {McLean},
  {Benford}, {Lonsdale}, {Blain}, {Mendez}, {Irace}, {Duval}, {Liu}, {Royer},
  {Heinrichsen}, {Howard}, {Shannon}, {Kendall}, {Walsh}, {Larsen}, {Cardon},
  {Schick}, {Schwalm}, {Abid}, {Fabinsky}, {Naes}, and {Tsai}}]{WISE10}
{Wright}, E.~L., {Eisenhardt}, P.~R.~M., {Mainzer}, A.~K., {Ressler}, M.~E.,
  {Cutri}, R.~M., {Jarrett}, T., {Kirkpatrick}, J.~D., {Padgett}, D.,
  {McMillan}, R.~S., {Skrutskie}, M., {Stanford}, S.~A., {Cohen}, M., {Walker},
  R.~G., {Mather}, J.~C., {Leisawitz}, D., {Gautier}, III, T.~N., {McLean}, I.,
  {Benford}, D., {Lonsdale}, C.~J., {Blain}, A., {Mendez}, B., {Irace}, W.~R.,
  {Duval}, V., {Liu}, F., {Royer}, D., {Heinrichsen}, I., {Howard}, J.,
  {Shannon}, M., {Kendall}, M., {Walsh}, A.~L., {Larsen}, M., {Cardon}, J.~G.,
  {Schick}, S., {Schwalm}, M., {Abid}, M., {Fabinsky}, B., {Naes}, L., {Tsai},
  C.-W., Dec. 2010. {The Wide-field Infrared Survey Explorer (WISE): Mission
  Description and Initial On-orbit Performance}. The Astronomical Journal 140,
  1868--1881.

\bibitem[{{Zappal\'a} et~al.(1990){Zappal\'a}, {Cellino}, {Farinella}, and
  {Kne\v zevi\'c}}]{zapetal90}
{Zappal\'a}, V., {Cellino}, A., {Farinella}, P., {Kne\v zevi\'c}, Z., 1990.
  {Asteroid families. I - Identification by hierarchical clustering and
  reliability assessment}. Astronomical Journal 100, 2030--2046.

\end{thebibliography}

\end{document}